\documentclass[12pt,preprint]{emulateapj}
\usepackage{apjfonts}
\usepackage{lscape}

\slugcomment{ApJ accepted version January 28, 2006}

\shorttitle{TFR in Cl0024 at z=0.4}
\shortauthors{Metevier et al.}

\begin{document}

\title{The Tully-Fisher Relation in Cluster Cl0024+1654 at z=0.4\altaffilmark{1} \altaffilmark{2}}

\altaffiltext{1}{Based on observations obtained at the W.\ M.\ Keck Observatory, which is operated jointly by the California Institute of Technology and the University of California.} 

\altaffiltext{2}{Based in part on observations with the NASA/ESA Hubble Space Telescope, obtained at the Space Telescope Science Institute, which is operated by the Association of Universities for Research in Astronomy, Inc., under NASA contract NAS 5-26555.}



\author{Anne J. Metevier\altaffilmark{3} and David C. Koo}

\altaffiltext{3}{NSF Astronomy and Astrophysics Postdoctoral Fellow}

\affil{University of California Observatories/Lick Observatory and Department of Astronomy and Astrophysics, University of California, Santa Cruz, CA 95064}

\email{anne@ucolick.org, koo@ucolick.org}

\author{Luc Simard}

\affil{Herzberg Institute of Astrophysics, National Research Council of Canada, Victoria, BC V9E 2E7, Canada}
\email{Luc.Simard@nrc-cnrc.gc.ca}

\and \vspace*{-0.3cm} \author{Andrew C. Phillips}

\affil{University of California Observatories/Lick Observatory, University of California, Santa Cruz, CA 95064 }
\email{phillips@ucolick.org}

\begin{abstract}

Using moderate-resolution Keck spectra, we have examined the velocity profiles of 15 members of cluster Cl0024+1654 at $z=0.4$.  WFPC2 images of the cluster members have been used to determine structural parameters, including disk sizes, orientations, and inclinations.  We compare two methods of optical rotation curve analysis for kinematic measurements.  Both methods take seeing, slit size and orientation, and instrumental effects into account and yield similar rotation velocity measurements.  Four of the galaxies in our sample exhibit unusual kinematic signatures, such as non-circular motions.  Our key result is that the Cl0024 galaxies are marginally underluminous ($0.50 \pm 0.23$ mag), given their rotation velocities, as compared to the local Tully-Fisher relation.  In this analysis, we assume no slope evolution, and take into account systematic differences between local and distant velocity and luminosity measurements.  Our result is particularly striking considering the Cl0024 members have very strong emission lines, and local galaxies with similar H$\alpha$ equivalent widths tend to be overluminous on the Tully-Fisher relation.  Cl0024 Tully-Fisher residuals appear to be correlated most strongly with galaxy rotation velocities, indicating a possible change in the slope of the Tully-Fisher relation.  However, we caution that this result may be strongly affected by magnitude selection and by the original slope assumed for the analysis.  Cl0024 residuals also depend weakly on color, emission line strength and extent, and photometric asymmetry.  In a comparison of stellar and gas motions in two Cl0024 members, we find no evidence for counter-rotating stars and gas, an expected signature of mergers.  

\end{abstract}


\section{Introduction}

The Tully-Fisher relation (TFR, Tully \& Fisher 1977) between disk galaxy luminosities and rotational velocities is an essential diagnostic of galaxy evolution.  By comparing galaxy evolution models to TFR observations at $z \sim 0$ through $z \sim 1$, it is possible to probe dark matter halo properties and galaxy star formation histories (e.g., Mo \& Mao 2000; Navarro \& Steinmetz 2000; Ferreras \& Silk 2001; Buchalter, Jimenez \& Kamionkowski 2001a, 2001b; Mo \& Mao 2004).  

Observations of the distant field galaxy TFR have clearly demonstrated luminosity evolution since $z \sim 1$, but the amount of evolution is contested.  Forbes et al.\ (1996), Vogt et al.\ (1996, 1997) and Rigopoulou et al.\ (2002) find $\le$1 mag brightening between $z \sim 0$ and $z \sim 1$, whereas Rix et al.\ (1997) find evidence for 1.5 mag brightening at $z \sim 0.25$.  In agreement with the latter work, Simard \& Pritchet (1998) and Mall\'{e}n-Ornelas et al.\ (1999) find evidence for 1.5--2 mag brightening at $z \sim 0.25 - 0.6$.  These two studies, however, also hint that there may be a change in TFR slope with redshift.

Recent studies using larger samples of field galaxies ($\geq$60 compared to sample sizes of $\la$20 in the studies mentioned above) by Ziegler et al.\ (2002) and B\"{o}hm et al.\ (2004) have provided further evidence for evolution in TFR slope, possibly reconciling previous findings.  More specifically, their work indicates that the mass-to-light ratios of the least massive galaxies evolve most strongly.  These authors attribute the differing results of previous studies to selection effects and small sample sizes.  Their results are consistent with kinematic studies of blue compact galaxies at $z \le 1$ (Koo et al.\ 1995, 1997; Phillips et al.\ 1997; Guzm\'{a}n et al.\ 1996, 1997, 1998).  The latter studies demonstrate that distant compact galaxies are overluminous as compared to the local TFR, in many cases an order of magnitude less massive than expected for their luminosities.

While the work done to date is promising, there are still ways in which distant Tully-Fisher studies can be improved.  For instance, selection effects are thought to be responsible for the differing results of some of the studies mentioned above.  However, while effects such as magnitude incompleteness have been explored in local studies (e.g. Willick 1994, Giovanelli et al.\ 1997), little work has yet been done toward quantifying and correcting for such effects on distant samples.  Furthermore, the velocity measurement techniques used in local and distant Tully-Fisher studies are often very different and may account for some of the discrepancies between previous results.  Authors of distant Tully-Fisher studies also tend to compare their samples to the relatively small local sample of Pierce \& Tully (1992) to draw evolutionary conclusions.  While this is a good starting point, larger and better-calibrated local samples now exist for comparison (e.g., Tully \& Pierce 2000; Kannappan, Fabricant \& Franx 2002).

We also note that while significant recent progress has been made investigating the evolution of the Tully-Fisher relation in the field, TFR evolution in the cluster environment is only beginning to be explored.  Many physical processes are expected to affect galaxies in this dense environment, including tidal encounters, interactions with the intracluster medium, and group infall.  All of these processes are predicted to affect cluster galaxy disks and gas, and in particular their mass-to-light ratios, thus affecting the TFR.

As a specific example, high-velocity tidal encounters between cluster galaxies and with the gravitational potential of the cluster itself (``harassment") have been modeled by Moore et al.\ (1996, 1998, 1999).  These encounters are expected to increase star formation rates while stripping galaxies of a significant fraction of their mass (Fujita 1998), thus decreasing their mass-to-light ratios.  Gnedin (2003) predicts that lower-density galaxies in clusters are more likely to be disrupted and lose a greater percentage of mass in tidal encounters than large spirals.   The mass-to-light ratios of low- and high-mass galaxies in clusters may therefore evolve differently.  In distant galaxy clusters, we have the opportunity to observe such evolutionary processes in action.  Distant cluster Tully-Fisher studies will therefore allow us to test model predictions and quantify evolutionary effects.

Few studies of the Tully-Fisher relation in distant clusters currently exist.  While studies of $z \sim 0.1$ clusters have been conducted with significant sample sizes ($>$50 galaxies; e.g., Rubin et al.\ 1999, Dale \& Uson 2003), kinematics of few $z>0.1$ cluster disks have been measured.  In a pioneering study, Vogt et al.\ (1993) measured the rotation velocities of two cluster galaxies at $z=0.20$ and $z=0.38$.  Franx (1993) also demonstrated rotation in an ``E+A" galaxy in Abell 665 at $z=0.18$.  All three objects were found to be $\leq 1$ mag brighter than local disk galaxies.  

More recently, Milvang-Jensen et al.\ (2003) analyzed the rotation curves of eight members of cluster MS1054-03 at $z=0.83$ and found that they were $\sim$1.5 mag more luminous than local field galaxies.  Bamford et al.\ (2005) have expanded this study with a larger sample size of 22 members of three clusters at $0.3 < z < 0.9$.  They have found that the cluster members are somewhat ($\sim$0.5 mag) more luminous than field galaxies at the same redshifts.  On the other hand, Ziegler et al.\ (2003; see J\"{a}ger et al.\ 2004 for details on analysis techniques), presented rotation curves of 13 members of three clusters at $0.3 < z < 0.5$ and found no evidence for luminosity evolution of cluster galaxies with respect to the field.  Other than these noteable exceptions, the majority of kinematic studies in distant clusters have focused on the fundamental plane of early-type galaxies.  These works indicate that the amount of luminosity evolution in cluster early-types is consistent with passive evolution out to $z = 0.83$ (van Dokkum \& Franx 1996; van Dokkum et al.\ 1998; Kelson et al.\ 1997, 2000a,b; J$\o$rgensen et al.\ 1999) and possibly out to $z = 1.27$ (van Dokkum \& Stanford 2003).  

As described above, environmentally-dependent physical processes are expected to affect cluster disks more strongly than early-types.  The distant cluster TFR may provide clues as to which processes are dominant.  As a first step, we have chosen to study the TFR in rich cluster Cl0024+1654 at $z=0.4$, where evolutionary processes are very likely to be apparent.   In general, morphological, luminosity, and color evolution have already been seen in cluster galaxies since $z \simeq 0.5$ (e.g., Dressler et al.\ 1997, J$\o$rgensen et al.\ 1999, and Butcher \& Oemler 1984, respectively).  The redshift of this cluster alone places it at an epoch where we would expect to observe evolutionary signatures.  Cl0024 was one of the two clusters originally studied by Butcher \& Oemler (1978) and thus has long been known to have a high fraction of blue, presumably star-forming members.  Furthermore, evidence has been found for evolutionary processes such as mergers (Lavery, Pierce, \& McClure 1992) and recent subcluster infall (Czoske et al.\ 2002).  

The structure of this paper runs as follows: in the next section, we discuss our observations of Cl0024+1654 and our sample selection for this study.  Note that our sample of 15 Cl0024 members is one of the largest samples of rotation curves presented thus far for galaxies within a single $z > 0.2$ cluster.  We detail our basic spectral and image reductions, two methods of rotation curve analysis, and we discuss the differences between local and distant galaxy velocity measurements in \S 3.  In \S 4, we present the Cl0024 Tully-Fisher relation, including a comparison to three local samples.  In \S 5, we discuss evolutionary indications from correlations between Tully-Fisher residuals and other galaxy properties.  We also take a look at the effect of magnitude incompleteness on our analysis.  We present our conclusions in \S 6.  Throughout this paper we assume $H_{0}=70$ km s$^{-1}$ Mpc$^{-1}$, $\Omega_{M} = 0.3$ and $\Omega_{\Lambda} = 0.7$.  Using this cosmology, one arcsecond corresponds to 5.332 kpc.  We present all photometry in the Vega system.

\section{Observations and Sample Selection}

Spectra of all galaxies presented in this paper were taken as part of a larger spectral survey of Cl0024+1654 (Metevier et al., in preparation).  Survey observations were made in September 1995, September 1997, and September 2001 with the Low Resolution Imaging Spectrograph (LRIS, Oke et al.\ 1995) on the Keck 10-m Telescopes.  All Cl0024 spectra were taken through multi-object slitmasks with slit widths of $0.9-1\arcsec$.  A variety of gratings (400, 600, 900, and 1200 lines/mm) and settings were used such that dispersions range from $0.65-1.85$\ \AA\ pixel$^{-1}$ (corresponding to $\sim$30-80 km s$^{-1}$ pixel$^{-1}$), and wavelength ranges cover at least H$\alpha$ (9155\AA, observed) through H$\beta$ (6780\AA, observed) and in some cases [\ion{O}{2}]  (5200\AA, observed) at the cluster redshift.  The spatial resolution of all spectra in our survey is 0.215$\arcsec$ pixel$^{-1}$.  Seeing ranged from $\sim 0.65-1\arcsec$ FWHM, and instrumental resolution ran from 1.0 to 2.9 \AA\ FWHM (40 to 120 km s$^{-1}$).  In many cases, slits were approximately aligned with galaxy major axes.  Slit misalignments have been explicitly included in our rotation curve analysis.

Archival Hubble Space Telescope (HST) images of Cl0024 were taken with WFPC2 as part of two programs.  Deep imaging of a single pointing at the center of the cluster was taken (proposal 5453, PI Turner) to study the mass of the cluster via strong lensing.  Exposure times are 25.2 ksec and 19.8 ksec in $B_{450}$ and $I_{814}$, respectively.  More recently, shallower (4.0-4.4 ksec) images of the outer regions of the cluster have been taken in $I_{814}$ (proposal 8559, PI Ellis) for a weak lensing study.  The latter program provides imaging of 39 non-contiguous pointings covering a circle of radius $\sim$15 arcmin, with a filling factor of approximately 50\%.  

Discrepancies between the results of Tully-Fisher studies of distant field galaxies have generally been attributed to different sample selections.  We have imposed very few selection criteria here:  we have restricted our Tully-Fisher sample to emission line cluster members with $0.37 < z < 0.41$, and we have required that the galaxies in this sample lie within the HST-imaged region of the cluster.  High-resolution WFPC2 images are essential to our structural analysis for galaxy disk sizes, inclinations, and position angles. In turn, these measurements are crucial for kinematic analysis.  We have also required disk inclination angles range between $30^{\circ} < i < 80^{\circ}$.  These simple selection criteria narrowed the original sample of 312 total objects in our survey to 44 cluster members eligible for Tully-Fisher analysis.  Note that we have not intentionally imposed selection based on colors, morphologies, or emission line strengths.  

Only 15 of the 44 objects in our Tully-Fisher-eligible sample have extended, high signal-to-noise emission lines such that we were able to measure a reliable terminal rotation velocity.  This low success rate imposes additional selection on top of the criteria described above.  We explore these selection effects further in Figure 1, where we compare the $R$ magnitudes, $B - R$ colors, morphological types (from Treu et al.\ 2003), bulge-to-total flux ratios (B/T), and disk scale lengths of galaxies in the Tully-Fisher and larger cluster samples.  In addition, in Figure 2, we present the distributions of emission line extents and projected cluster-centric distances for the galaxies in the final Tully-Fisher sample (15 galaxies) and the Tully-Fisher-eligible sample (44 galaxies).  Emission line extents were measured in the spatial direction along each galaxy's spectrum from the top to the bottom of the galaxy's emission, as some galaxies do not exhibit symmetric emission about the galaxy centroid.  We estimate that the errors on the extent measurements are on the order of one pixel ($\sim 0.215\arcsec$).  In Tables 1 and 2, we quantify the mean values, standard deviations, minimum and maximum values of the compared properties for the different galaxy samples.\notetoeditor{Please place Figures 1 and 2 and Tables 1 and 2 here.}

In general, we find that the galaxies in our Tully-Fisher sample are relatively bright and blue as compared to the other emission-line cluster members and the total cluster sample.  Furthermore, they tend to be large, late-type, disk-dominated systems.  Galaxies in the final Tully-Fisher and Tully-Fisher-eligible samples are similarly distributed within the cluster. However, galaxies in the final sample have the largest emission line extents.  We are therefore mindful of the fact that Cl0024 members with suppressed or centrally concentrated star formation, perhaps due to physical processes within the cluster, are likely not included in our Tully-Fisher study.  In Table 3, we give basic data for the final sample of 15 Cl0024 members with reliable rotation curve measurements.\notetoeditor{Please place Table 3 here.}  

\section{Data Reduction and Analysis}

\subsection{Spectral and Image Reduction}

We reduced all LRIS spectra using Python modules developed by D.\ Kelson (see Kelson et al.\ 2000, Kelson 2003).  To prepare the spectra for this reduction, we performed overscan subtraction using standard IRAF procedures, and we cosmic ray reduced the frames with a task written by coauthor L.\ Simard.  Using Python modules, each spectrum and dome flat was rectified in the dispersion direction to account for ``pincushion"-shaped distortion introduced by LRIS optics.  We then combined dome flat fields; dome flats from two of the three runs were taken with the same grating angles and telescope elevation and azimuth as were used during the object exposures in order to reproduce and combat the effects of fringing.  The resulting combined flats were used to correct for pixel-to-pixel variations in two dimensions.  Along the length of the slit, the flats were also used to correct for instrumental distortions and variations in slit width.  We then rectified the spectra in the spatial direction and wavelength calibrated against arc lamp exposures and known night sky lines.  We sky-subtracted the resulting spectra, then co-added multiple exposures, accounting for sub-pixel offsets in the spatial direction.  Finally, we made redshift measurements using cross-correlation software developed by coauthor A.\ C.\ Phillips.

We obtained pipeline-processed WFPC2 images from the Canadian Astronomy Data Centre archive service.  Remaining reduction consisted of cosmic ray removal and structural modeling of galaxies in the images.  To remove cosmic rays, we combined exposures of the cluster core, which were taken with no offsets.  Exposures of the outer regions of the cluster had subpixel offsets and were therefore individually treated using the LACOSMIC routine (van Dokkum 2001).  We fit a two-dimensional deVaucouleurs bulge $+$ exponential disk model to each galaxy using GIM2D (Simard et al.\ 2002).  This analysis provided galaxy bulge-to-total flux ratios (B/Ts) and disk inclinations, scale lengths, and position angles.  These measurements are noted in Table 3.

\subsection{Rotation Curve Analysis}

Measuring the rotation velocities of distant galaxies from optical spectra is a challenging undertaking which requires comparison of emission line data to emission line models.  In part this is because by $z \sim 0.4$, the size of a typical galaxy ($\sim$1$\arcsec$) is roughly equivalent to the size of the seeing disk and the slit width.  Therefore, light and velocity information from a substantial fraction of a galaxy are gathered in its spectrum.  The resulting rotation curve is far from the optimal narrow slice down the galaxy major axis acquired in long slit studies of nearby galaxies.  Below we provide a more complete list of factors one must consider when converting raw emission line measurements into maximum disk rotation velocities for distant galaxies:

\begin{enumerate}
\item Spatial distribution of observable galaxy emission
\item Velocity distribution of observable galaxy emission
\item Inclination of the disk
\item Seeing blurring of the observed velocity distribution
\item Instrumental blurring of observed emission due to instrument optics
\item Slit width with respect to the size of the galaxy
\item Slit position angle with respect to the galaxy semi-major axis
\item Slit position with respect to the galaxy center 
\item Anamorphic de-magnification in the spectral direction
\item Pixelization of the observed emission line
\item Atmospheric dispersion
\item Thickness of the disk
\item Presence of bulge, bar, warps
\end{enumerate}

Previous studies of distant galaxies have mainly used one of two techniques for deriving velocities from rotation curves.  These two techniques are 1) comparison of velocity measurements from Gaussian fits to observed and model emission lines at several spatial locations (Vogt et al.\ 1996, 1997; Rigopoulou et al.\ 2002; Ziegler et al.\ 2002, 2003; B\"{o}hm et al.\ 2004; J\"{a}ger et al.\ 2004), and 2) direct comparison of two-dimensional information from observed and simulated emission lines (Simard \& Pritchet 1998, 1999; Milvang-Jensen et al.\ 2003; Bamford et al.\ 2005).  We have implemented both analysis methods in our study of Cl0024 members; below we refer to our version of method 1 as GAUSS2D and our implementation of method 2 as GELFIT2D.

Both of our analysis techniques rely upon the same emission line modeling routine.  The routine we have adopted assumes an infinitely thin, round disk galaxy with no bulge, bar, or warps.  This model galaxy has a truncated exponential intensity profile and a symmetric ``step function" (e.g., Persic \& Salucci 1991) or an arctangent (e.g., Willick 1999) velocity distribution.  We found that both velocity functions yielded similar rotation velocity measurements  (generally within 10 km s$^{-1}$) for our sample.  We therefore have chosen to use the arctangent function for this paper:
\begin{equation}
V(r) = \frac{2V_{arc}}{\pi}{\rm arctan}(\frac{r}{r_{to}})
\end{equation}

\noindent where $V_{arc}$ is the asymptotic rotation velocity, and $r_{to}$ is the ``turnover" radius which defines the slope of the inner velocity curve.  We note that more elaborate empirical descriptions of galaxy velocity distributions may better fit observed rotation curves (see, e.g., Courteau's 1997 study of nearby galaxies).  They also introduce new parameters to be modeled and therefore we neglect them here.

Our emission line modeling routine first samples the assumed intensity and velocity distributions according to a grid size specified by the user.  In order to sample at least 10 times across each galaxy, we generally subsampled above the resolution of our model data according to the following prescription, where {\it pixscale} is the spatial pixel scale of our data, $r_{d}$ is the disk scale length, and $i$ is the galaxy's inclination:
\begin{equation}
{\rm subsampling\ factor} = \frac{{\it pixscale} * 5}{r_{d} * {\rm cos}(i)}
\end{equation}

\noindent The routine then convolves the model galaxy's light and velocity profiles with a circular Gaussian seeing disk with a given FWHM.  The seeing-degraded model galaxy is masked with a slit, taking into account the slit width as well as the position angle difference between the slit and galaxy major axis.  The subpixel position of the galaxy center with respect to the center of the slit can also be accounted for (as in GELFIT2D, \S 3.2.2) by varying the position of the model galaxy with respect to the model slit mask.  The model galaxy is then compressed in the spatial direction due to the instrument anamorphic factor and further convolved with a circular Gaussian representing blurring from instrumental optics.  The resulting light and velocity profiles are rebinned and resampled to create a model emission line matching the resolution of the user's data.  

In short, while our emission line modeling routine makes some simplifying assumptions about the intensity and velocity distributions of emission within a galaxy (items 1 and 2 on our list of modeling considerations above), it fully accounts for factors 3 through 10.  Through the implementation discussed in \S 3.2.3, atmospheric dispersion (factor 11) can also be accounted for.  We note that disk thickness (factor 12) has not been incorporated in rotation curve analysis techniques presented previously in the literature, nor is it considered here.  Similarly, the effects of a bulge, bar, and warps have not been incorporated in distant galaxy rotation curve analysis.  Another improvement that could be made in the future is the direct use of color images to model emission line intensity distribution, rather than the assumption of an exponential distribution.

\subsubsection{GAUSS2D}

Our GAUSS2D velocity analysis follows the method used by Vogt et al.\ (1996, 1997), Rigopoulou et al.\ (2002), Ziegler et al.\ (2002, 2003), B\"{o}hm et al.\ (2004), and J\"{a}ger et al.\ (2004).  We determined raw velocity measurements by fitting Gaussian profiles to the strongest observed emission lines at each row along the slit.  These fits provided Gaussian amplitudes, centers, widths, and linear background measurements, with errors on each of the measurements.  Fits were required to meet certain criteria chosen to throw out noise peaks or measurements with unreliable S/N. These criteria were: Gaussian amplitude $> 10$ counts, center $> 2\sigma$ in significance, and 1\AA \ $<$ FWHM $<$ 20\AA.  Gaussian centers were converted to velocities and were required to differ no more than 100 km s$^{-1}$ from one of the neighboring measurements.  Generally, several emission lines were fit for each object (double Gaussians were fit to [\ion{O}{2}]), and the resulting velocity information was converted to a single weighted mean and error at each row along the slit.  
 
To determine the rotation velocities of the galaxies in our sample, we applied the same Gaussian fitting to emission lines modeled with a grid of $V_{arc}$ and $r_{to}$ values.  Grids were initially implemented with $V_{arc}$ ranging between 50 and 450 km s$^{-1}$ in 50 km s$^{-1}$ increments, and with $r_{to}$ ranging between 0.1 and 0.5\arcsec\ in 0.1\arcsec\ increments.  We compared observed and model velocity distributions to create corresponding $\chi^{2}$ grids, then interpolated to determine the optimal $V_{arc}$ and $r_{to}$ measurements (corresponding to the minimum $\chi^{2}$) with 68\% confidence intervals.  This process was then repeated with more refined grids for each galaxy such that $V_{arc}$ values were tested in increments of 10 km s$^{-1}$, and $r_{to}$ in increments of 0.05\arcsec.  For simplicity, we assumed model galaxies were centered in the slit when using this method.

In comparison with previous implementations of this analysis method, our version incorporates some improvements.  The emission line models we use fully take into account factors 3 -- 7 and 9 -- 10 above; it is unclear whether previous authors have considered anamorphic magnification (9) or pixelization (10) in their models.  Furthermore, GAUSS2D allows the option of simultaneously fitting multiple emission lines for a given galaxy.  Optimization of the measured velocity and confidence intervals are determined using a $\chi^{2}$ grid.  We estimated further contributions to errors by varying the following four parameters: disk inclination and PA were each varied by $\pm10^{\circ}$, the seeing FWHM by $\pm0.2\arcsec$, and the instrumental profile FWHM by $\pm0.25$ pixels.  We added the largest range of errors determined by varying these four parameters in quadrature with the formal fitting errors on our original measurements to derive a final set of errors on our $V_{arc}$ and $r_{to}$ measurements.  We note that previous studies (e.g., Vogt 1996, 1997) have only taken into account errors due to estimation of the inclination and position angle, beyond the formal fitting errors.

\subsubsection{GELFIT2D}

Our second rotation curve analysis method is very similar to the ELFIT2D analysis described in detail by Simard \& Pritchet (1999) and used by Simard \& Pritchet (1998), Milvang-Jensen et al.\ (2003), and Bamford et al.\ (2005).  A Metropolis optimization wrapper was applied to our emission modeling routine in order to directly compare observed and modeled emission lines, vary modeled parameters and determine best-fit values.  (See below for a more detailed description of this optimization method.)  

Specifically, this routine takes the following inputs: 1) a two-dimensional sky- and continuum-subtracted galaxy emission line ``thumbnail image"; 2) an initial guess and range of possible values for each modeled parameter; 3) a list of fixed parameter values.  Fixed parameters include the type of velocity field to be used in the emission line models (arctangent or step function), seeing FWHM, observed wavelength of the emission line, an optional second observed wavelength (e.g.\ to be used in modeling the [\ion{O}{2}] doublet), galaxy inclination, position angle between the slit and galaxy major axis, slit width, dispersion, spatial pixel scale, instrumental FWHM, a subsampling factor to account for pixelization, a sigma value describing the noise in the thumbnail image background, and the CCD gain. 

GELFIT2D then optimizes the following eight parameters: $V_{arc}$; $r_{to}$; $r_{em}$, the emission line scale length; $r_{hole}$, the inner truncation radius (assuming a central hole in emission, e.g.\ due to dust extinction); offsets of the galaxy center with respect to the input thumbnail image of emisson along and across the slit (can be subpixel); and the offset of the slit with respect to the center of the thumbnail in the dispersion direction.  In the case of [OII] emission, the flux ratio between the two lines in the observed doublet is also modeled.

To start the model, an initial grid of parameters within the user-specified range of possible values is first created.  The corresponding emission line models are compared to the input data image by calculation of the likelihood function for each model.  The most likely initial model is then chosen as a starting point for optimization.  The Metropolis algorithm (Metropolis et al.\ 1953, Saha \& Williams 1994) is used to refine the best-fit parameter values.  This algorithm generates random perturbations about the initial model parameter values.  The size of these perturbations is controlled by the Metropolis ``temperature": if a model is hot, the perturbations are large; if the model is cold, small perturbations are tried.  At each iteration, the likelihood that the new set of parameter values is correct, P$_{1}$, is calculated.  This likelihood is compared to that of the previous iteration, P$_{0}$.  The new trial is adopted if P$_{1}>$ P$_{0}$.  If on the other hand P$_{1}<$ P$_{0}$, the new trial is still adopted P$_{1}$/P$_{0}$ of the time.  This allows the algorithm to walk itself out of local $\chi^{2}$ minima, a particularly important feature for low signal-to-noise data.  Furthermore, the shallower the minimum, the higher the probability that the model will walk itself out.  

Convergence is achieved when the difference between two likelihood values separated by 100 iterations is less than $3\sigma$ of the likelihood value fluctuations.  Once the Metropolis algorithm converges, GELFIT2D determines 68\% confidence intervals by Monte-Carlo sampling the parameter space where the likelihood has been maximized.  In other words, GELFIT2D does not assume that confidence intervals are Gaussian.  This is again important in the case of low signal-to-noise data.  The optimization algorithm employed in GELFIT2D has also been used in the galaxy structural modeling routine GIM2D; see Simard et al.\ (2002) for more details.  GELFIT2D outputs a log file specifying best-fit parameter values with errors as well as a thumbnail image of the best-fit model emission line and a residual (data -- best-fit model) image.  Since GELFIT2D does not allow for simultaneous fitting of multiple emission lines, we fit several lines separately for each galaxy, then took the weighted mean as our final GELFIT2D velocity measurement.  We determined contributions to the errors on our $V_{arc}$ and $r_{to}$ measurements by varying disk inclinations and PAs, as well as the seeing and instrumental FWHMs, and adding the largest range of resulting errors in quadrature with the formal weighted errors from our original measurements, as described at the end of \S 3.2.1.

Our implementation of GELFIT2D incorporates some improvements over the similar ELFIT2D routine.  Specifically, GELFIT2D accounts for considerations 3 -- 10 on our modeling list, and provides some further flexibility in our assumptions regarding consideration 1 by allowing for a central hole in the emission intensity distribution.  GELFIT2D, unlike ELFIT2D, explicitly takes into account the position angle offset between the slit and galaxy major axis as well as possible offsets of the slit center with respect to the galaxy centroid.    However, we note that ELFIT2D does have some advantages over GELFIT2D, the most noteable being the option to specify an empirical seeing PSF and instrumental profile rather than assuming circular Gaussian profiles.

\subsubsection{Comparison of analysis methods}

Images and two-dimensional emission lines, as well as output from both of our rotation curve analysis techniques are shown in Figure 3.  GAUSS2D and GELFIT2D share the same emission line modeling routine and are therefore not completely independent.  However, they differ in two significant ways: 1) the routines use different methods for comparing observed and model emission lines, and 2) the optimization routines used to converge on a best-fit model are different.  Both routines have advantages and disadvantages that may affect our measurements, and we list them here.  

Our GAUSS2D fitting routine has two distinct advantages.  For one, it depends little on any assumption about the spatial light profile of a galaxy in the determination of the galaxy's rotation velocity.  In this routine, velocity measurements are made separately for each spatial row; the velocity at each row is determined from the center of a Gaussian fit to the galaxy's emission along the dispersion direction.  For the most part, the relative amplitudes of the Gaussian peaks in each spatial row do not affect the relative velocity measurements.  Due to seeing blurring and the small pixel scale of our data, emission seen in adjacent spatial rows is not entirely independent, though.  A galaxy's light profile could therefore have a small effect on the resulting velocity measurements.  

A second advantage of this routine is that it concatenates velocity measurements from a user-specified number of emission lines for a given object before determining $V_{arc}$.  That is, the emission lines can be fit together or separately.  A disadvantage of this method is that it assumes galaxy emission profiles are Gaussian along the dispersion direction.  This does not necessarily reflect the true emission distribution and may bias velocity measurements, as they are derived from the Gaussian-fit centers of the emission.  Furthermore, this routine simply does not take full advantage of the two dimensions of information available from a spectrum.\notetoeditor{Please place Figure 3 here.}

GELFIT2D is a much simpler fitting routine in that it directly compares data to emission line models, without any intermediate (e.g., Gaussian fitting) steps.  GELFIT2D fits a larger number of parameters than the GAUSS2D routine, providing additional measurements.  These include the offset of a galaxy within a slit, the radius within which emission is truncated at the center of a galaxy, the emission line scale length, and the flux ratio of the [\ion{O}{2}] doublet.  Furthermore, GELFIT2D has a much more sophisticated method for parameter optimization than GAUSS2D.  While GELFIT2D only runs on one emission line at a time, a possible benefit is independent modeling of the position of the galaxy within the slit at red and blue wavelengths, indirectly allowing us to account for atmospheric dispersion (factor 11).  It should be noted that GELFIT2D also has disadvantages, the main one being the assumption of a symmetric exponential intensity profile.  This assumption can strongly affect resulting velocity measurements, particularly for galaxies with asymmetries, spiral arms, and/or star-forming hot spots.  

Both routines take on average 15-20 minutes to run per galaxy.  GAUSS2D requires real-time user input of $V_{arc}$ and $r_{to}$ ranges and increments to be tested.  GELFIT2D is completely automated, but optimizes eight parameters rather than two.  Because GELFIT2D fits each emission line separately, this routine may take longer if a large number of lines are fit.  Run-times for both GAUSS2D and GELFIT2D are also dependent upon the amount of subsampling desired: highly inclined galaxies will take the longest to run because the subsampling factor will be largest.

For ease of comparison, in Table 4 we again list the modeling factors described in \S 3.2.  We also describe how and whether we have taken them into account using GAUSS2D and GELFIT2D.  In Table 5, we list velocity measurements for our sample of 15 Cl0024 members determined via both routines.  In general, we find that the two methods yield gratifyingly similar results, as can be seen in Figure 4.  We note that in all cases, emission was measured out to radii beyond 2.2$r_{d}$, the radius at which local galaxy studies often determine rotation velocities (e.g., Courteau 1997).  Objects with truncated emission were not eliminated directly because of the small extent of their emission, but because the turnovers in their velocity distributions were not strong enough for us to measure $V_{arc}$.  Of course, the latter criterion does bias our sample toward galaxies with large emission line extents (see Figure 2).  However, we note that some galaxies with relatively large emission line extents did not make it into our final Tully Fisher sample because the signal-to-noise in the outskirts of their emission was too low for us to reliably measure $V_{arc}$.\notetoeditor{Please place Figure 4 and Tables 4 and 5 here.}

\subsubsection{Unusual Galaxies in the Sample}

Four of the objects in our sample yielded particularly odd emission distributions and/or poor kinematic fits (see the Appendix for more notes on individual objects).  One object (TFR 01) exhibits high-ionization emission lines characteristic of an AGN (e.g., [\ion{Ne}{5}] $\lambda3426$) and has a measured [\ion{O}{2}] flux ratio I($\lambda3729$)/I($\lambda3726$) = 0.4, indicating electron densities over 1000 cm$^{-3}$.  This emission is particularly concentrated toward the center of the galaxy: we measure a disk scale length of 1.1$\arcsec$ and an emission scale length of only $0.2\arcsec$.  We caution that the derived rotation velocity may represent motion of gas in the galaxy core only.    

The other three objects are Butcher-Oemler ``blue" galaxies whose morphologies were examined by Lavery, Pierce, \& McClure (1992; hereafter LPM92) and linewidths were measured by Koo et al.\ (1997).  These are the only three objects in our sample that overlap with the LPM92 and Koo et al.\ studies.  One of these objects (TFR 10) appears to contain a large, knotty ring of emission.  LPM92 describe this galaxy as having a ``disturbed" morphology.  This is the largest and most rapidly rotating galaxy in our sample, and its velocity curve appears to turn over at large radii.  The morphological and kinematic distortion in this galaxy indicate that it may have undergone a recent interaction.  Koo et al.\ report a linewidth $\sigma = 120$ km s$^{-1}$ for this object, where here $\sigma$ refers to the standard deviation measured by fitting a Gaussian to the spatially collapsed (one-dimensional) emission.  We measure a Gaussian $\sigma = 115$ km s$^{-1}$ for TFR 10 but find $V_{arc} = 370$ km s$^{-1}$.  We discuss possible sources for the differences between these measurements below.

The remaining two ``unusual" objects (TFR 04 and TFR 08) have odd morphologies poorly fit with our GIM2D bulge + disk modeling.  Both objects have such steeply rising rotation curves at their cores that a pure step function ($r_{to} \simeq 0\arcsec$) velocity distribution convolved with realistic seeing could not reproduce the observed velocity curve.  Therefore, our velocity measurements appear to be overestimated.  This is most clearly apparent in output from GAUSS2D (see Figure 3).  Since the GELFIT2D velocity measurements closely match those from GAUSS2D, we assume they are overestimated as well.  As in the case of TFR 10, linewidths measured by Koo et al.\ (1997) are significantly smaller (40 and 50 km s$^{-1}$ for TFR 04 and TFR 08, respectively) than our $V_{arc}$ measurements (120 and 135 km s$^{-1}$).  However, we measure similar Gaussian linewidths of 75 and 40 km s$^{-1}$ for the two objects, generally in agreement with Koo et al.  Both of these objects were flagged as probable mergers/interacting galaxies by LPM92; we consider them tidally distorted.  

In the three cases where our samples overlap (TFR 04, TFR 08, and TFR 10), our $V_{arc}$ measurements are much larger than the velocity widths measured by Koo et al.\ (1997) from spatially collapsed emission.  This is due in part to inherent differences between the measurement techniques.  Emission toward the outer regions of a galaxy (where velocities are often highest) is given more weight in rotation curve analysis than in linewidth analysis, where the emission signature is collapsed into one dimension.  Furthermore, Koo et al.\ fit the linewidths of the three galaxies in question with Gaussians despite the fact that the velocity profiles of these galaxies are complex, exhibiting 2-3 local peaks.  However, we again caution that our V$_{arc}$ measurements may be overestimated for these galaxies, as TFR 10 exhibits a turnover at large radii not accounted for by our models, and TFR 04 and TFR 08 show possible non-circular motions.

It is not surprising that a significant fraction of the galaxies in our Cl0024 Tully-Fisher sample ($27\%$) exhibit unusual kinematics.  Nor is it surprising that these galaxies were relatively poorly fit by our rotation curve analysis routines.  Simulations of the rotation curves of interacting galaxies (Barton, Bromley, \& Geller 1999) predict complicated bumps and wiggles, and observations of close pairs in the nearby field (Barton et al.\ 2001) reveal a significant fraction ($\sim10\%$) with centrally concentrated emission.  In the nearby Virgo cluster, $\sim50\%$ of the 89 galaxies studied by Rubin, Waterman, \& Kenney (1999) exhibit kinematic disturbances.  While the scales of many of these disturbances are too small for us to observe at $z=0.4$, some general trends (such as velocity turnovers at large radii) can be apparent (as in TFR 10).  In their study of $z \sim 0.4$ cluster galaxies, Ziegler et al.\ (2003) note that four of the 30 disk galaxies they observed exhibited peculiar kinematics and were difficult to model, so they were not included in the authors' Tully-Fisher analysis.

Anomalous kinematic signatures have been found in distant field galaxy studies as well.  Rix et al.\ (1997) identify $\sim$20$\%$ of their $z \sim 0.25$ sample as kinematically unusual, with low [\ion{O}{2}] flux ratios.  Furthermore, Simard \& Pritchet (1998) found that $\sim$25$\%$ of their sample showed unusual kinematics, generally with emission line scale lengths smaller than disk scale lengths.

\subsection{Comparing Velocity Width Measurements}

In the majority of literature on the distant galaxy TFR (including our own early analysis of the Cl0024 TFR, Metevier \& Koo 2004), authors directly compare the luminosities and velocities of galaxies in their distant samples to a locally-derived TFR without addressing differences in velocity measurement methods for distant and local galaxies.  However, systematic differences do exist between different measurement methods.  These have been discussed in several TFR studies at $z \lesssim 0.25$ (e.g., Courteau 1997; Raychaudhury et al.\ 1997; Rix et al.\ 1997; Barton et al.\ 2001; Kannappan, Fabricant, \& Franx 2002).  In this section, we describe local velocity measurement methods and we detail our procedure for converting our $V_{arc}$ measurements to velocity widths that can be directly compared to locally-derived values.

In \S 4, we compare the Tully-Fisher relation in Cl0024 to the TFR derived from three local galaxy samples: those studied by Pierce \& Tully (1992, hereafter PT92); Tully \& Pierce (2000, hereafter TP00); and Kannappan, Fabricant, \& Franx (2002, hereafter KFF02).  Both PT92 and TP00 derive velocity widths from 21cm observations.  For each galaxy, they measure $W_{20}$, the line width at $20\%$ of the peak intensity.  From this they derive an inclination-corrected ``rotation width'' $W_{R}^{i}$ by dividing by sin $i$ and applying the turbulence correction given in Tully \& Fouqu\'{e} (1985).  

KFF02 measure ``probable min-max'' rotation velocities from optical emission line observations.  The ``probable min-max'' velocity, $V_{\mathrm{pmm}}$, is half the difference between {\it (a)} the velocity determined to statistically exceed $90\%$ of the velocities in the galaxy's rotation curve and {\it (b)} the velocity determined to statistically exceed only $10\%$ of the velocities in the rotation curve.  Both KFF02 and Raychaudhury et al.\ (1997) demonstrate that $V_{\mathrm{pmm}}$ is a robust measurement of a galaxy's velocity, in part because it does not rely on a particular rotation curve shape.  However, this method is difficult to apply to distant galaxies, for which one must model seeing, slit, and instrumental effects and therefore make some assumptions about the shape of a galaxy's intrinsic velocity distribution.

In order to directly compare to other local TFR studies, KFF02 convert their $V_{\rm{pmm}}$ measurements to velocity widths equivalent to $W_{50}$, radio line widths measured at $50\%$ of peak intensity.  KFF02 derive a relationship between $V_{\rm{pmm}}$ and $W_{50}$ by directly comparing optical and radio observations of the galaxies in their sample.  They also find that $W_{50} = W_{20} - 20$ km s$^{-1}$ (see also Haynes et al.\ 1999).  KFF02 adopt the nomemclature $W_{V_{\rm{pmm}}}$ to describe their $W_{50}$-equivalent velocity measurements derived from $V_{\rm{pmm}}$.

In Table 6, we list the simple steps we have taken to convert our $V_{arc}$ measurements to velocity widths equivalent to $W_{R}^{i}$ and $W_{V_{\mathrm{pmm}}}$ for comparison to PT92, TP00, and KFF02.  We note that in step 1, we have taken the inclination correction out of our $V_{arc}$ measurements.  Instead we inclination-correct the measurements in later steps in order to keep the order of our procedure consistent with the methods used in the local studies.  In step 2, we have calibrated the conversion between $V_{arc}$ and $V_{\rm pmm}$ by analyzing data in Courteau (1997), in which $V_{arc}$ and $V_{\rm pmm}$ measurements were made for the same sample of local galaxies.  Using an iterative least-squares fit to Courteau's data, we find:
\begin{equation}
2V_{arc} = 0.80(\pm 0.02)V_{\rm pmm} + 23(\pm 10) {\rm \ km \ s}^{-1}
\end{equation}

\noindent In Table 7, we list our original $V_{arc}$ measurements for each Cl0024 cluster member, as well as our derived $W_{R}^{i}$ and $W_{V_{\rm pmm}}$ values.  \notetoeditor{Please place Table 6 here.}  We note that we have rounded all velocity measurements to the nearest 5 km s$^{-1}$ to reflect the precision of our data.

\subsection{Galaxy Luminosities}

We measured observed total $R$ magnitudes and 3$\arcsec$ diameter aperture $B-R$ colors from LRIS imaging of the Cl0024 field (600 s exposures in both $B$ and $R$).  This photometry was then corrected for Galactic extinction in two ways in order to compare our measurements directly to those used in the literature.  One method was an application of the extinction correction given by Burstein \& Heiles (1982, hereafter BH82), a correction of 0.09 mag in $R$ in the Cl0024 field, and a correction of 0.15 mag in $B$.  The other method was application of the correction from Schlegel, Finkbeiner, \& Davis (1998, hereafter SFD98), 0.15 mag in $R$.

The observed photometry was converted to restframe total $M_{B}$ magnitudes and aperture $(U - B)_{0}$ colors by C.\ N.\ A.\ Willmer.  The procedure for doing this is described in detail in Willmer et al.\ (2005) and basically follows these steps: empirical template spectra from Kinney et al.\ (1996) are redshifted to the cluster $z$, and the resulting SEDs are convolved with the known transmission curves of our filters, then compared with the observed photometry.  Restframe photometry was calculated from the best-fitting template spectrum for each galaxy.

The resulting restframe $M_{B}$ magnitudes were then corrected for internal extinction using the relations given in either Tully \& Fouqu\'{e} (1985, hereafter TF85) or in Tully et al.\ (1998, hereafter T98), depending on the literature we compared to.  We note that the internal extinction corrections given in TF85 are inclination-dependent, whereas the corrections given in T98 are depend both on a galaxy's inclination and intrinsic luminosity (or rotation velocity, $W_{R}^{i}$).  

In all, we derived four sets of restframe $M_{B}$ magnitudes for comparison to the literature.  (1) $M_{B}^{\rm dist}$ magnitudes were derived using SFD98 Galactic extinction corrections and the TF85 internal extinction corrections.  This method follows that used in recent {\it distant} TFR studies such as Ziegler et al.\ (2002, 2003), Milvang-Jensen et al.\ (2003), and B\"{o}hm et al.\ (2004). (2) $M_{B}^{PT92}$ magnitudes were derived using BH82 Galactic extinction corrections and the TF85 internal extinction corrections, for comparison to Pierce \& Tully (1992).  (3) $M_{B}^{\rm TP00}$ magnitudes were derived using SFD98 Galactic extinction corrections and the T98 internal extinction corrections, for comparison to Tully \& Pierce (2000).  (4) $M_{B}^{\rm KFF02}$ magnitudes were derived using BH82 Galactic extinction corrections and the T98 internal extinction corrections, following the method used by Kannappan, Fabricant, \& Franx (2002).  In this latter case, we calculated $W_{R}^{i}$ to be applied in the T98 internal extinction correction using the method outlined in Table 6, but with steps 6 and 7 switched.  In other words, the turbulence correction was applied to the rotation velocity before the inclination correction in this case only.  This same procedure was adopted by KFF02.

The internal extinction corrections we have derived for each galaxy are noted in Table 8; absolute $M_{B}$ magnitudes and restframe $(U - B)_{0}$ colors are in Table 7.  Restframe colors have been calculated using the BH82 Galactic extinction corrections (see above) and T98 internal extinction corrections for comparison to KFF02.  Like KFF02, we use an extrapolated internal extinction correction for the $U$ band (see equation 3 in their paper).\notetoeditor{Please place Table 7 and 8 here.}

\section{The Cl0024 Tully-Fisher Relation}

We present the $B$ band Tully-Fisher relation for our sample of 15 Cl0024 members in Figure 5.  In each of the four panels of this figure, we make a different comparison of the Cl0024 TFR to a local Tully-Fisher relation.  This was motivated by the following: in some of the very first studies of the distant galaxy TFR, Vogt et al.\ (1996, 1997) compared their results to the local relation of PT92.  Since then, larger local samples (such as TP00 and KFF02) have been better analyzed, for example with improved internal extinction corrections, and with somewhat different local TFR results.  However, the authors of more recent studies of the distant galaxy TFR (e.g., Milvang-Jensen et al.\ 2003, Ziegler et al.\ 2003, B\"{o}hm et al.\ 2004) have continued to compare to the local PT92 relation so that their results can be more directly juxtaposed against those of Vogt et al.  In Figure 5, we compare to PT92 for the same reasons, but we also compare to the local relations of TP00 and KFF02 in order to be more comprehensive in our analysis.\notetoeditor{Please place Figure 5 here.}

In each panel of Figure 5, the inverse Tully-Fisher relation (velocity as a function of luminosity) is shown: in magnitude-limited samples, fitting the inverse relation helps to avoid biases due to asymmetric faint-end scatter in luminosities (e.g., Schechter 1980).  In Figure 5(a), we present a ``raw" comparison between the Cl0024 TFR and the PT92 relation.  By this we mean that we directly compare our $V_{arc}$-$M_{B}^{\rm dist}$ relation to their $W_{R}^{i}$-$M_{B}^{\rm PT92}$ relation without accounting for differences in velocity measurement techniques or for differences in extinction corrections used to calculate restframe $B$ magnitudes.  This is the type of comparison commonly made in distant galaxy TFR studies in order to deduce luminosity evolution results.

Because our sample spans a limited range in luminosities, we examine the Cl0024 TFR in Figure 5(a) assuming the slope determined by PT92.  Using the biweight method (Beers, Flynn, \& Gebhardt 1990), we determine an offset from the PT92 field relation of $<$$\Delta$TF$>$ $= -0.69 \pm 0.24$ mag.  The value of this offset remains consistent to within $\sim$0.1 mag whether we use the total Cl0024 TFR sample or only the objects with ``normal" kinematics.  We note that PT92 measured a $B$ band zeropoint offset between local cluster and field samples, indicating that local cluster galaxies are $\sim 0.25$ mag less luminous in this band than galaxies in the nearby field.  This would make our Cl0024 sample $0.94$ mag brighter than the PT92 local cluster sample.

However, this result may be misleading.  If we convert our $V_{arc}$ measurements for Cl0024 members to $W_{R}^{i}$ (see \S 3.3) and derive restframe $M_{B}$ using the same Galactic and internal extinction corrections as those used by PT92 (see \S 3.4), we can make a ``matched", rather than ``raw", comparison of the distant cluster TFR to the local relation.  This ``matched'' comparison is shown in Figure 5(b).  Recalculating the biweight offset of the Cl0024 relation from PT92, we find that Cl0024 members are in this case only $0.16 \pm 0.23$ mag brighter on average than local galaxies.  In other words, matching our velocity and luminosity measurement techniques to those used by PT92 has changed our results by $\sim$0.5 mag.

We have also made ``matched" comparisons to the more recently-published local TF relations of TP00 and KFF02, shown in Figure 5(c) and 5(d).  The TP00 sample is larger than PT92, drawn from a range of environments, and internal extinction corrections have been improved in this study.  KFF02 is also an improvement upon PT92.  Their sample comes from the Nearby Field Galaxy Survey (NGFS, Jansen \& Kannappan 2001), a sample of local galaxies chosen without preference for morphology, color, size, environment, or any other galaxy property.  Specifically, in Figure 5(d) we compare to the subsample of Sa-Sd galaxies in KFF02 with $M_{R} < -18$ and inclinations $i > 40^{\circ}$.  We note that we find a slight offset ($\sim$0.2 mag) between the TP00 and KFF02 local relations.  A similar offset has been found independently by Kannappan (private communication; see also KFF02), confirming that our velocity and luminosity measurement ``matching'' is accurate.  

If we assume uniform shifts from the local TF relations with no change in slope or dispersion, it is clear from Figure 5 that the Cl0024 sample appears marginally {\it underluminous} as compared to both TP00 ($<$$\Delta$TF$>$ $= 0.29 \pm 0.17$ mag) and KFF02 ($<$$\Delta$TF$>$ $= 0.50 \pm 0.23$ mag; we note that the zeropoint errors on the reference TP00 and KFF02 Tully-Fisher relations are $\lesssim$0.1 mag).  This is somewhat surprising, as almost all other distant TFR studies have found that by $z \sim 0.25$, galaxies are somewhat overluminous as compared to the local TFR (see the introduction of this paper for a list of examples).  However, our results are consistent with PT92 who, as noted above, found that local cluster galaxies are $\sim$0.25 mag less luminous in the B band than their local field counterparts.  Furthermore, Kannappan et al.\ (2003) found a slight underluminosity when they re-analyzed the distant field sample of Vogt et al.\ (1997).  Specifically, Kannappan et al.\ found that this $<$$z$$>\ \sim 0.5$ sample is $\lesssim 0.5$ mag underluminous as compared to galaxies in the NGFS with similar rotation velocities and similarly large emission line equivalent widths.  Kannappan et al.\ suggest that this may be due to a decrease in stellar mass fraction with lookback time.  Our results not only hint at such an underluminosity, but also highlight the importance of the choice of comparison sample for studies of Tully-Fisher evolution, and of ``matching" measurements in the resulting comparison.

As in distant field galaxy TFR studies, there has thus far been some evidence for brighter cluster galaxies, relative to a given rotation velocity, in the past.  Milvang-Jensen et al.\ (2003) studied 19 field spirals at $0.15 < z < 0.90$ and eight members of cluster MS1054-03 at $z = 0.83$.  By making a ``raw" comparison of their field sample to the local PT92 relation, they determined a relation between galaxy redshifts and their offsets from the local TFR: $\Delta$TF $\approx$ $-1.6z$, assuming $H_{0}=75$ km s$^{-1}$ Mpc$^{-1}$ and $q_{0}=0.05$.  Converting to our cosmology, they predict $\Delta$TF $=-0.93$ mag at $z=0.4$.  Our ``raw" Cl0024 TFR, with $\Delta$TF $=-0.69$ mag, is slightly less luminous than their prediction.

However, Milvang-Jensen et al.\ showed that their cluster sample is somewhat ($\sim$0.5--1.0 mag) brighter than their field sample at the same redshift.  They suggest that this luminosity enhancement may reflect increased star formation in spirals falling into cluster MS1054 for the first time.  More recently, Bamford et al.\ (2005) have expanded this study with larger field (58 galaxies) and cluster (22 galaxies) samples and have derived similar results.  These results appear to contradict ours, as well as those of Ziegler et al.\ (2003), who studied the TFR for 13 cluster galaxies at $z \sim 0.5$ and found no significant luminosity evolution.  This may reflect true differences between the spiral populations of different clusters and/or differences between our sample selection techniques.  In the future, we intend to explore this apparent discrepancy with a larger sample of several hundred spirals spanning a range of environments in the GOODS fields and Extended Groth Strip.  Spectroscopy in these fields is underway as part of the DEEP2 Survey (Davis et al.\ 2003) and the Team Keck Treasury Redshift Survey (Wirth et al.\ 2004).

\section{Discussion of Tully-Fisher Residuals}

From Figure 5, one can see that the observed scatter in the Cl0024 TFR spans the 3$\sigma$ confidence intervals of the PT92 and TP00 local relations.  Furthermore, scatter in the Cl0024 TFR is comparable to that found by KFF02, whose sample is unusual in that it has not been ``pruned'' to include only normal spirals.  Increased scatter in the Tully-Fisher relation has been linked on the observational side to galaxy-galaxy interactions (Barton et al.\ 2001).  On the theoretical side, it has been linked to the ratio of dark to luminous mass within galaxies of different luminosities (Salucci, Frenk, \& Persic 1993).  Using the biweight technique and comparing to the KFF02 relation, we find $\sigma_{TF} = 1.00$ mag in the Cl0024 TFR (objects with ``normal" kinematics only, $\sigma_{TF}$ increases a bit to 1.14 mag if we include the full sample).  This is comparable to but slightly larger than the 0.82 mag dispersion found by KFF02.  After accounting for photometry errors (0.15 mag) and errors on our velocity measurements (corresponding to 0.42 mag), we determine the intrinsic Cl0024 $\sigma_{TF} = 0.90$ mag.  This is somewhat larger than the intrinsic scatter ($\sigma_{TF} = 0.57$ mag) measured by KFF02 in the $B$ band, and is much larger than the total scatter, including measurement errors, reported by PT92 ($\sigma_{TF} = 0.41$ mag) and by TP00 ($\sigma_{TF} = 0.38$ mag).  Below, we analyze this scatter, or ``residuals'' from the Cl0024 Tully-Fisher relation, in more detail.

Before presenting this analysis, we note that authors of local Tully-Fisher studies (e.g., Courteau 1997, Willick 1999) have found that using $V_{2.2}$ instead of $V_{arc}$ reduces scatter in the TFR.  Here, $V_{2.2}$ is the rotation velocity measured at 2.15$r_{d}$, where $r_{d}$ is the disk scale length; the radius 2.15$r_{d}$ corresponds to the point at which the rotation amplitude of a pure exponential disk reaches its maximum (e.g., Freeman 1970).  As a test, we have redetermined the Cl0024 TFR by calculating $V_{2.2}$ for the cluster members, entering the measured values of $V_{arc}$, $r_{to}$, and 2.15$r_{d}$ for each galaxy into equation (1) of this paper (see \S 3.2 for the equation and Tables 3 and 5 for the measurements).  Furthermore, we have redetermined radio-equivalent velocity widths based on the $V_{2.2}$ measurements, replacing step 2 of Table 6 with a similar calibration derived from the min-max velocities, arctangent velocities, and turnover radii presented in Courteau (1997), and the disk scale lengths given in Courteau (1996).  We find that using $V_{2.2}$ does reduce the total scatter in the Cl0024 TFR, measured as a simple offset from the KFF02 relation, to 0.42 mag; this is comparable to our measurement errors alone.  However, the choice of $V_{arc}$ or $V_{2.2}$ does not qualitatively change the results of our Tully-Fisher analysis or our analysis of Tully-Fisher residuals in Cl0024.  Therefore, we continue with our TF residual analysis based on $V_{arc}$ measurements.

\subsection{TF Residuals and Other Fundamental Galaxy Properties}

Several authors (e.g., Verheijen 2001) have searched for a ``third parameter" to the Tully-Fisher relation for spiral galaxies, analogous to the dependence of velocity dispersion and surface brightness on galaxy size for early-types (known as the fundamental plane, Djorgovski \& Davis 1987).  Until recently, no strong evidence has been given for a relationship between Tully-Fisher residuals ($\Delta$TF) and other galaxy properties.  However, in their study of nearby field galaxies, KFF02 have shown that $\Delta$TF and galaxy colors are correlated, reflecting an influence from galaxy star formation histories.  This finding is supported by a similar trend found in the nearby Ursa Major cluster by Verheijen (2001).  Although the number of galaxies in our sample is relatively small, we attempt here to look for correlations between Tully-Fisher residuals and other fundamental galaxy properties in Cl0024, in the hopes that they may provide clues about evolutionary processes in the cluster.  

In the following analysis, we define $\Delta$TF as the difference between a galaxy's observed luminosity and its predicted luminosity, derived from a comparison of its rotation velocity to the Cl0024 TFR (characterized by a simple 0.50 mag offset from the KFF02 relation):
\begin{equation}
\Delta{\rm TF{\it (mag)}} = M_{B}^{\rm KFF02} - \{^{-}19.83 - 10.09[{\rm \ log}(W_{V_{\rm pmm}}) -2.5] + 0.50\}
\end{equation}

\noindent We have determined errors on these TF residuals by adding our photometry errors (0.15 mag) in quadrature with our velocity measurement errors, converted to magnitudes.  Our $\Delta$TF measurements are listed in Table 9.\notetoeditor{Please place Table 9 here.}  

We discuss TF residuals in terms of magnitudes in this paper in order to compare to local Tully-Fisher residual analysis (e.g., KFF02).  However, we note that this can be problematic for distant Tully-Fisher studies, including our own, which have significant luminosity selection effects.  In future studies, a better treatment of the data will be to discuss {\it velocity} residuals from the Tully-Fisher relation for distant galaxies.  Therefore, in Table 9 we also list the Cl0024 member TF residuals as velocities, determined by comparing the galaxies' luminosities to the Cl0024 TFR:
\begin{equation}
\Delta{\rm TF{\it (km\ s^{-1})}} = W_{V_{\rm pmm}} - 10^{\{[M_{B}^{\rm KFF02} + 19.83 - 0.50]/^{-}10.09 + 2.5\}}
\end{equation}

\noindent Again, we add our photometry errors (converted to velocities) and velocity errors in quadrature to determine the errors on the residuals.  We also round the velocity residuals to the nearest 5 km s$^{-1}$ to reflect the precision of our data.

\subsubsection{Luminosity}

In Figure 6 (top panels), we compare TF residuals ($\Delta$TF) to the fundamental parameters of the TFR itself: luminosity ($M_{B}^{\rm KFF02}$) and velocity (log $W_{V_{\rm pmm}}$).  As can be seen in Figure 6(a), there is no dependence of $\Delta$TF on luminosity: using the Spearman rank test on the ``normal'' sample, we find a 77\% probability that the correlation between $\Delta$TF and $M_{B}^{\rm KFF02}$ occurred by chance.  A formal line fit to this sample yields a slope of $-0.42 \pm 0.27$, indicating that the data are inconsistent with the null hypothesis (zero slope, or no correlation) at only the 1.6$\sigma$ level.  We shade the left side of Figure 6(a) to indicate our selection bias against galaxies with $M_{B} > -19.5$.  Only one galaxy fainter than this limit (TFR 08) has made it into our sample due to its very strong emission.\notetoeditor{Please place Figure 6 here.}

\subsubsection{Velocity}

On the other hand, in Figure 6(b), TF residuals for our sample appear to be correlated with rotation velocity (the chance probability of the correlation is six in 10,000), such that the most slowly rotating systems are most overluminous.  Furthermore, our formal fit to the ``normal'' sample is inconsistent with the null hypothesis at the 9$\sigma$ level.  This may indicate slope evolution of the TFR and be evidence for mass-segregated luminosity evolution, a possible outcome of harassment (e.g., Gnedin 2003).   Indeed, if we do not assume a local Tully-Fisher slope in Cl0024, we find that an inverse fit to the ``normal'' sample shown in Figure 5(d) yields a Tully-Fisher slope of $-7.04\pm 0.30$ as opposed to the KFF02 slope of $-10.09\pm0.39$.  Evidence for a slope change in the same direction has also been observed in the field (Simard \& Pritchet 1998, Mallen-Ornelas et al.\ 1999, Ziegler et al.\ 2002, B\"{o}hm et al.\ 2004), so this may not be specific to the cluster environment.  

Our sample is small and incomplete at low luminosities, so we explore how magnitude selection affects measurements of slope evolution before placing great weight on our result.  While incompleteness effects on Tully-Fisher measurements have been studied in great detail at low redshifts (e.g., Willick 1994, Giovanelli et al.\ 1997), these effects are significant yet relatively unexplored in distant Tully-Fisher studies.  We have shaded Figure 6(b) to demonstrate how luminosity incompleteness at $M_{B} > -19.5$ affects the distribution of data points in this diagram.  It is possible that the correlation between velocity and $\Delta$TF we find may be artificially tight because we cannot observe galaxies to the lower left in this plot.  We investigate this further in Figure 7, where we demonstrate how magnitude selection affects Tully-Fisher measurements of three randomly generated galaxy populations: a low-dispersion sample (black points), a medium-dispersion sample (dark grey points), and a high-dispersion sample (light grey points).  We have perturbed the velocities and luminosities of the galaxies in each of these samples around the $B$-band TFR measured in KFF02, increasing the perturbations as the galaxies' luminosities and velocities decrease to reflect the fact that we typically incur larger measurement errors at the low-luminosity end of the Tully-Fisher relation.  The numbers of galaxies we have generated in each luminosity bin follow the shape of the $g$-band luminosity function given in Blanton et al.\ (2001), with 0.45 subtracted from the $g$-band magnitude, as this is the typical $g$-$B$ color of an Sab galaxy (Fukugita et al.\ 1995).\notetoeditor{Please place Figure 7 here.}

In Figure 7(a), we show the Tully-Fisher relation for the three samples we have generated.  This panel is shaded to demonstrate a selection bias against galaxies with $M_{B} > -19.5$, as seen in the Cl0024 sample.  For a given low velocity, one can see that only the highest-luminosity data points make it into the unshaded region of the diagram.  In Figure 7(b), we present the corresponding log(velocity)-$\Delta$TF relation, also shaded to reflect our magnitude selection bias.  The low-velocity, high-luminosity galaxies show up as a tail toward the upper left of the unshaded region of the diagram.  This tail is particularly prominent for the high-dispersion sample, and could lead one to measure Tully-Fisher slope evolution despite the fact that no slope change was input into the sample.  This is clear in Table 10, where we list the the input Tully-Fisher and log(velocity)-$\Delta$TF relations, as well as the measured Tully-Fisher and log(velocity)-$\Delta$TF relations for the galaxies with $M_{B} \leq -19.5$ in each of the three generated samples.\notetoeditor{Please place Table 10 here.}

The magnitude selection bias results in increasingly ``shallower'' measured Tully-Fisher slopes with increasing dispersion.  (We note that ``shallower'' Tully-Fisher slopes actually appear steeper in our diagrams, as we have plotted the inverse Tully-Fisher relation in Figures 5 and 7.)  This slope change is in the same direction as that indicated by our own data and measured in other distant Tully-Fisher studies (e.g., Ziegler et al.\ 2002, B\"{o}hm et al.\ 2004).  An apparent slope change in this direction due to magnitude incompleteness has also been found in the local studies of Willick (1994) and Giovanelli et al.\ (1997).  In our case, the selection bias results in an increasingly prominent and steep correlation between velocity and $\Delta$TF with increasing dispersion.  From Figure 7 and Table 10, we see that any Tully-Fisher slope evolution inferred particularly from our low-luminosity Cl0024 data points is strongly affected by magnitude incompleteness in combination with measurement errors and the intrinsic dispersion of the galaxies' properties.  However, if there were no slope evolution, we would expect to see a turn-off to $\Delta$TF = 0 at the high-velocity end in the log(velocity)-$\Delta$TF diagram (see Figure 7b).  We do not see this in the Cl0024 sample (Figure 6b).  

It is possible that an incorrect measurement of the slope of the Tully-Fisher relation would cause us not to see a turn-off to $\Delta$TF = 0 at the high-velocity end of the Figure 6(b).  This would be the case particularly if we measured a ``steeper'', or more strongly negative, slope than that which characterized the true distribution of our sample.  However, as explained above, it is much more likely that we would measure an overly ``shallow'' slope, given our magnitude incompleteness.  In the bottom panels of Figure 7, we demonstrate the effect of a ``shallow'' slope measurement on the log(velocity)-$\Delta$TF relation.  In Figure 7(c), we plot the log(velocity)-$\Delta$TF relation for the high-dispersion sample of generated data, as this is the sample for which it is most difficult to measure a Tully-Fisher slope.  Here, as in Figure 7(b), we have derived $\Delta$TF by comparing the generated galaxy magnitudes to their expected magnitudes, given their velocities and the {\it known input} Tully-Fisher relation.  In Figure 7(d), we again plot log(velocity) versus $\Delta$TF, but this time we have calculated $\Delta$TF by comparing generated galaxy magnitudes to those expected from the {\it measured} Tully-Fisher relation (see Table 10), assuming the input relation was unknown.  As can be seen in this diagram, an underestimated (or less strongly negative) Tully-Fisher slope still results in a turn-off, or here a turn-up, at the high-velocity end of the log(velocity)-$\Delta$TF relation.  Again, we do not see this in the Cl0024 sample (see Figure 6b), and we therefore conclude that the Tully-Fisher slope evolution we see in the cluster may be real.  

We would need a larger galaxy sample to discern more conclusively whether or not the slope of the Tully-Fisher relation is evolving.  Furthermore, quantifying the effect of magnitude selection on our measurement of Tully-Fisher slope evolution requires Monte-Carlo simulations with knowledge of galaxy number counts built in.  That work is beyond the scope of this paper.  We caution here that the apparent slope evolution found by other authors (e.g., B\"{o}hm et al.\ 2004) may also be affected by magnitude incompleteness.  It is also possible that their results are driven by low-luminosity galaxies with truncated rotation curves or other kinematic anomalies that have resulted in underestimated rotation velocities (Kannappan \& Barton 2004).  This does not appear to be the case for our sample (see \S 5.1.6), as Cl0024 members with the smallest emission line extents are generally the most underluminous.  Again, this points toward the fact that the Tully-Fisher slope evolution we see in Cl0024 may be real.  Future studies with larger sample sizes will clarify this result.

\subsubsection{Color}

In Figure 6(c), we probe the correlation between Cl0024 $\Delta$TF and galaxy colors.  We find a marginally significant correlation, with a chance probability of 5\%.  The formal line fit to the ``normal'' sample is inconsistent with the null hypothesis at the 5$\sigma$ level.  As previously mentioned, a relationship between TF residuals and color has recently been found in the nearby field (using $B-R$ colors) by KFF02 and in the Ursa Major cluster (using $B-I$ colors) by Verheijen (2001).  However, Kannappan \& Barton (2004) stress that local galaxies with kinematic anomalies tend to be outliers from the color-$\Delta$TF relation.  Indeed, we find a much less significant correlation (chance probability of 19\%) if we fit the color-$\Delta$TF relation for our total Cl0024 sample, including four ``anomalous'' galaxies, rather than the ``normal'' sample.  KFF02 infer that the color-$\Delta$TF relationship reflects the galaxies' star formation histories, as predicted by Bell \& de Jong (2001).  For comparison, we plot the KFF02 relation, converted to our restframe passbands (S.\ J.\ Kannappan, private communication), in Figure 6(c) and find that it lies very close to our data points.  

We also stress here that $M_{B}$, $W_{V_{\rm pmm}}$, and $(U-B)_{0}$ are all inter-related.  It is well-known at lower redshifts that galaxy velocities and colors are correlated such that less massive galaxies have bluer colors.  In the top panel of Figure 8, we show that the same is true in Cl0024.  If we follow the example of Rix et al.\ (1997), we can fit $M_{B}$, log $W_{V_{\rm pmm}}$, and $(U-B)_{0}$ simultaneously according to the following equation:
\begin{equation}
{\rm log}(W_{V_{\rm pmm}}) = \alpha + \beta(M_{B}) + \gamma(U-B)_{0}
\end{equation}

\noindent The fit to our total sample yields $\alpha = 1.28$, $\beta = -0.065 \pm 0.35$, and $\gamma = 0.27 \pm 1.42$.  This is in good agreement (though we have much larger errors) with the simultanous fit of Rix et al., who found $\alpha = 1.82$, $\beta = -0.07 \pm 0.08$, and $\gamma = 0.28 \pm 0.25$ for $z \sim 0.25$ field galaxies.  Note that Rix et al.\ used $(B-R)$ colors in their fit rather than $(U-B)_{0}$; $(B-R)$ roughly corresponds to $(U-B)_{0}$ at $z=0.4$, neglecting $k$-corrections.\notetoeditor{Please place Figure 8 here.}

In the bottom panel of Figure 8, we further address the relationship between $M_{B}$, $W_{V_{\rm pmm}}$, and $(U-B)_{0}$.  Here, we have re-plotted the color-$\Delta$TF relation for Cl0024, where the solid line in this panel represents a fit to the ``normal'' data points.  If we compare the color-velocity relation for Cl0024 members (top panel of Figure 8) to the velocity-$\Delta$TF relation (see Figure 6), we can re-determine the color-$\Delta$TF relation in Cl0024.  The dotted line in the lower panel of Figure 8 shows the color-$\Delta$TF relation derived in this way.  Clearly, the latter relation fits the data points well, suggesting that in the case of Cl0024, the color-$\Delta$TF relation may not be fundamental, but instead a natural outcome of the color-velocity and velocity-$\Delta$TF relations.

\subsubsection{Emission Line Strength}

Closely tied to the color-$\Delta$TF relation is the correlation between $\Delta$TF and global H$\alpha$ equivalent width (EW) found by KFF02.  While $(U-B)_{0}$ colors trace the ratio between young and somewhat older stellar populations in galaxies, EW(H$\alpha$) traces very recent star formation, on timescales of Myr.  In Figure 6(d), we explore the relationship between global EW(H$\alpha$) and $\Delta$TF.  Here, we have measured global EW(H$\alpha$) from integrated one-dimensional (1D) spectra using SPLOT in IRAF.  For each galaxy, the integrated 1D spectrum was extracted by summing the two-dimensional spectrum along the full spatial extent of the galaxy's emission.  This is the same procedure that was used to extract integrated 1D spectra and determine the global EW(H$\alpha$) of the galaxies presented in KFF02 (see Jansen et al.\ 2000).  Because the line strengths measured for both the local and distant samples have been made across each entire galaxy, we do not expect aperture effects to affect our results.  We again point out that are able to detect emission to a radius beyond two scale lengths for each of the galaxies in our sample.  Furthermore, we do not expect surface brightness dimming to play a role, as this would affect our ability to detect a galaxy's emission and the continuum used to normalize our equivalent width measurements in the same way.

We find a relatively strong EW(H$\alpha$)-$\Delta$TF correlation in Cl0024 (chance probability of 4\% for the ``normal'' sample).  This is driven, however, by the strongest emission line object in our sample.  Excluding this object from our analysis, we find that the correlation essentially goes away: we find a chance probability of 14\% and a slope consistent with the null hypothesis.  As in the case of the color-$\Delta$TF relation, it is possible that the lack of a strong EW(H$\alpha$)-$\Delta$TF relation in Cl0024 indicates that there are galaxies with anomalous kinematic signatures within our sample. 

We also include Figure 6(d) in our analysis to address the concern (e.g., in Kannappan et al.\ 2003) that distant Tully-Fisher studies are based on galaxy samples with very strong emission lines as compared to local samples.  This may affect our interpretation of Tully-Fisher evolution, as we would expect galaxies with large negative EWs to be somewhat overluminous according to the EW(H$\alpha$)-$\Delta$TF correlation found by KFF02 and seen in Figure 6.  As noted earlier, Kannappan et al.\ re-analyzed the Vogt et al.\ (1997) sample and found that this distant sample is actually somewhat {\it underluminous} as compared to local galaxies with similarly large EW(H$\alpha$).

Our sample is comprised of most of the strongest emission line galaxies in Cl0024.  While cluster galaxies generally do not have strong emission lines as compared to galaxies in the field, our sample does have somewhat stronger lines than the KFF02 sample: the majority of the galaxies in our sample have $-80 < $ EW(H$\alpha$) $ < -15$, whereas the majority of the KFF02 sample lies in the range $-55 < $ EW(H$\alpha$) $ < 0$.  This offset can be seen in Figure 6(d), where we compare a line fit to our ``normal'' sample (solid line) to the relationship between EW(H$\alpha$) and $\Delta$TF found by KFF02 (see also Kannappan 2001 where the calibration is given in the $B$ band).  Due to their strong emission, we would expect the galaxies in our sample to be overluminous, given their rotation velocities, as compared to the KFF02 sample.  The fact that we instead find slight evidence for an underluminosity is therefore all the more striking.  

\subsubsection{Disk Size}

Courteau \& Rix (1999) postulate that a correlation between $\Delta$TF and galaxy sizes may reflect the relative contributions of stellar disks and dark matter halos to total galaxy masses.  However, these authors find little evidence for such a correlation.  We find essentially no evidence for a relationship between the two parameters (42\% chance probability for the ``normal'' sample, inconsistent with null hypothesis at only the 3$\sigma$ level), as can be seen in Figure 6(e).  In similar investigations, only marginal correlations between $\Delta$TF and galaxy surface brightnesses have been found by KFF02 and Verheijen (2001).  The lack of a strong correlation found between $\Delta$TF and disk sizes could be influenced by the fact that it is difficult to measure rotation curves for small disk systems.  Certainly, our Cl0024 sample is comprised of relatively large disk galaxies, as discussed in \S 2.

\subsubsection{Emission Line Extent}

In several previous studies (including Vogt et al.\ 1996, Rix et al.\ 1997, Simard \& Pritchet 1998, Kannappan \& Barton 2004, and Bamford et al.\ 2005), the concern has been raised that galaxies with truncated rotation curves may appear overluminous on the Tully-Fisher diagram because of the difficulty of measuring their maximum rotation velocities.  This may particularly affect distant, low-luminosity galaxies, for which the outer regions of optical emission lines, where velocities are highest, are very faint.  Kannappan \& Barton suggest that distant TFR studies (e.g., Ziegler et al.\ 2002, B\"{o}hm et al.\ 2004) which report a Tully-Fisher slope change may be affected by this problem.

In Figure 6(f), we test whether truncated rotation curves may be affecting our results.  There we plot Cl0024 member emission line extents against their Tully-Fisher residuals.  We find a marginally significant correlation (chance probability of 5\%, but the ``normal'' sample is inconsistent with the null hypothesis at only the 2.6$\sigma$ level) such that galaxies with low emission line extents are most underluminous.  This is the opposite of what we would expect to find if emission line truncation was affecting our Tully-Fisher analysis.  However, we note that our sample does not span a wide range of emission line extents.  Furthermore, our sample is not complete at low luminosities, and therefore we do not place great significance on our finding of a Tully-Fisher slope change (see \S 5.1.2). 

\ 

The strongest relationship shown in Figure 6 is the dependence of $\Delta$TF on log($W_{V_{\rm pmm}}$) in Cl0024, which may be affected by magnitude incompleteness.  While we do not make strong evolutionary claims based on this finding, we still address a concern expressed by Barton et al.\ (2001) that a poor determination of the TFR slope and/or sample selection effects can result in false correlations of galaxy characteristics with TF residuals.  Based on this reasoning, Barton et al.\ essentially negate a relationship between galaxy asymmetries and $\Delta$TF claimed by Zaritsky \& Rix (1997).  In \S 5.1.2, we explored the effects of magnitude selection on the relationship between velocity and $\Delta$TF, and we presented evidence that our data suggest a real change in Tully-Fisher slope.  

To further test the strength of the correlation we have measured, we have re-analyzed the dependence of $\Delta$TF on log($W_{V_{\rm pmm}}$) for the ``normal'' Cl0024 sample, varying the slope of the TFR significantly.  We find that the chance probability of the velocity-$\Delta$TF correlation consistently remains below a few percent for a wide range of Tully-Fisher slopes, from $-4.0$ to $-16.0$.  The slope of the KFF02 relation is $-10.09$, and the slope we derive from an inverse Tully-Fisher fit to the ``normal'' sample is $-7.04$.  However, the correlation breaks down for Tully-Fisher slopes around $-3.0$ and $-2.0$, and we note that an unweighted forward fit to the ``normal'' sample yields a slope of $-2.7$.  As mentioned in \S 4, an inverse Tully-Fisher fit, with velocity treated as the dependent variable, is preferred over a forward fit for magnitude-limited samples such as ours.  Still, we caution that the strength of the velocity-$\Delta$TF correlation we find in Cl0024 may be affected by the Tully-Fisher slope we have chosen for the analysis.

\subsection{Probing Direct Evidence for Interactions}

Earlier in \S 5, we reported evidence for an increase in scatter in the Cl0024 TFR as compared to local field samples that have been ``pruned'' to include only morphologically normal spirals.  It is possible that an increase in scatter could be due to a high incidence of interactions within the cluster.  Supporting evidence comes from Barton et al.\ (2001), who found that the Tully-Fisher relation for a sample of 90 nearby spirals in close pairs shows 25\% more scatter than the reference local TFR.  In the nearby Virgo cluster, Rubin, Waterman \& Kenney (1999) find a high fraction (50\%) of disk galaxies with kinematic distortions; such distortions are predicted by Barton, Bromley \& Geller (1999) for interacting galaxies.  Cl0024 is known to contain merger candidates (Lavery, Pierce \& McClure 1992), and in \S 3.2.4 we have detailed four cluster members in our sample of 15 which have unusual kinematic distributions.   Furthermore, we find little evidence for a strong relationship between galaxy colors and Tully-Fisher residuals in Cl0024.  Kannappan \& Barton (2004) find that interacting and kinematically anomalous galaxies tend to be outliers on this relation, and therefore our result may indicate a higher fraction of anomalous galaxies in our sample than is apparent upon inspection of our (relatively low-resolution) data.

Likely signatures of recent mergers and tidal interactions include structural anomalies.  We therefore examine Cl0024 TF residuals versus photometric residual and asymmetry measurements from GIM2D modeling (Figure 9, top panels).  Photometric residual and asymmetry parameters were measured at 2$r_{1/2}$ according to the method described in Tran et al.\ (2001, 2003; see also Simard et al.\ 2002 and Schade et al.\ 1995 for descriptions of the asymmetry measurement).  Figure 9 demonstrates a very weak correlation between TF residuals and GIM2D residuals (35\% probability of a chance correlation for the ``normal'' sample, line fit is inconsistent with the null hypothesis at only the 3$\sigma$ level), but a slightly stronger one (15\% chance probability for the ``normal'' sample, line fit inconsistent with null hypothesis at the 4$\sigma$ level) between $\Delta$TF and photometric asymmetries.  The latter result lends weak support to the findings of Zaritsky (1995) and Zaritsky \& Rix (1997), who demonstrate that nearby asymmetric spirals tend to be overluminous as compared to the local TFR.  However, we again note that Barton et al.\ (2001) argue that the results presented by Zaritsky \& Rix can be explained by an incorrect Tully-Fisher slope measurement.  Because our Cl0024 sample is small and does not cover a large dynamic range in luminosity, we have adopted a local Tully-Fisher slope.  Therefore, Barton et al.'s concern may also apply to our result.\notetoeditor{Please place Figure 9 here.}

In the bottom panels of Figure 9, we test possible relationships between TF residuals and galaxy emission line residuals and asymmetries.  (For ease of comparison, we provide $\Delta$TF as well as photometric residuals and asymmetries in Table 9.)  Barton et al.\ (2001) have found that kinematic distortions are good predictors of $\Delta$TF in a study of close pairs of galaxies in the field.  Furthermore, the lack of a strong correlation between galaxy colors and $\Delta$TF in Cl0024 suggests that our Tully-Fisher sample may contain a substantial fraction of kinematically anomalous galaxies (see \S 5.1.3 for further discussion).  In a local cluster, however, Dale \& Uson (2000) find minimal trends between $\Delta$TF and rotation curve shapes and asymmetries.  Furthermore, Dale et al.\ (2001) and Dale \& Uson (2003) find little correlation between rotation curve shapes and asymmetries and galaxy clustercentric distances.  We have measured emission line residuals and asymmetries from GELFIT2D output residual images according to the following equations:

\begin{equation}
{\rm GEL\ residual} = \frac{\sum_{i,j} \frac{1}{2}| R_{i,j} + R_{i,j}^{180}| } {\sum_{i,j} I_{i,j}} - \frac{\sum_{i,j} \frac{1}{2}| B_{i,j} + B_{i,j}^{180}| } {\sum_{i,j} I_{i,j}}
\end{equation}

\begin{equation}
{\rm GEL\ asymm} = \frac{\sum_{i,j} \frac{1}{2}| R_{i,j} - R_{i,j}^{180}| } {\sum_{i,j} I_{i,j}} - \frac{\sum_{i,j} \frac{1}{2}| B_{i,j} - B_{i,j}^{180}| } {\sum_{i,j} I_{i,j}}
\end{equation}

\noindent where $I_{i,j}$ is the intensity of pixel $(i,j)$  in the original 2D emission line image, $R_{i,j}$ is the intensity of pixel $(i,j)$  in the residual emission line image, and $B_{i,j}$ is an estimate of the background intensity of pixel $(i,j)$.  Superscripts reflect the fact that these measurements are made with respect to 180$^{\circ}$ rotations.  In Figure 9, one can see that there are essentially no correlations between $\Delta$TF and GELFIT2D residuals and asymmetries in Cl0024 (line fits to ``normal'' sample are consistent with the null hypothesis).  This would seem to support the results of Dale et al.\ described above.  However, our sample (15 galaxies) is small, as was the original sample of Dale \& Uson (2000, 14 galaxies), so we caution that our conclusion may be affected by small number statistics.  Furthermore, our kinematic data were taken at a much lower resolution than typical data for local galaxies, and our data are also affected by seeing blurring and instrumental effects.  Therefore, it is not clear that our GELFIT2D residual and asymmetry measurements are very sensitive to intrinsic kinematic distortions in Cl0024 members.

As a last probe of interactions in Cl0024, we note that counter-rotating galaxies have been found in nearby field samples (e.g., by KFF02) and are a predicted outcome of galaxy mergers (Rubin 1994).  In Figure 10, we check for counter-rotation of gas and stars in two members of our Cl0024 Tully-Fisher sample for which we have high enough S/N continuum data to make stellar velocity measurements.  We believe that this is the first direct comparison of stellar and gas rotation for such distant galaxies.  One of the galaxies highlighted in Figure 10 (TFR 01) is a likely AGN; the other (TFR 04) appears to be tidally distorted (see discussion in \S 3.2.4).  We find no evidence for counter-rotation in these galaxies.  However, we note that the majority of nearby counter-rotators are early-type galaxies (e.g., Kuijken, Fisher, \& Merrifield 1996; Morse et al.\ 1998).  These early-types may represent merger remnants, whereas at least a subsample of Cl0024 emitters may be in the beginning stages of interactions or undergoing a different class of interactions, such as harassment.\notetoeditor{Please place Figure 10 here. Thanks!}

\section{Conclusions and Future Work}

In this paper, we have measured the rotation velocities of 15 members of galaxy cluster Cl0024 at $z=0.4$, and we have presented one of the first Tully-Fisher analyses for a distant cluster.  Our key points are the following: 

\begin{enumerate}
\item We have measured Cl0024 member velocities via two different rotation curve analysis methods: GAUSS2D, which closely resembles the method used by Vogt et al.\ (1996, 1997), and GELFIT2D, which is similar to the method used by Simard \& Pritchet (1998, 1999).  Both methods yield similar rotation velocity measurements, within the errors. 
\item We find that four of the galaxies in our sample have unusual kinematic properties or emission distributions.  One of these galaxies is a likely AGN, another has a knotty ring of star formation.  The remaining two have odd morphologies and are poorly fit by our velocity models, implying possible non-circular motions and/or recent interactions.  Similar fractions ($\sim25\%$) of kinematically ``anomalous" galaxies have been found in the distant field studies of Rix et al.\ (1997) and Simard \& Pritchet (1998).  In the local Virgo cluster, Rubin, Waterman, \& Kenney (1999) find that as many as 50\% of spirals exhibit unual kinematic distributions.
\item The Cl0024 Tully-Fisher relation is consistent with no luminosity evolution, or perhaps a marginal {\it underluminosity}, as compared to the local sample of Kannappan, Fabricant, \& Franx (2002).  In our analysis of the Cl0024 TFR, we have demonstrated the importance of accounting for differences between distant and local galaxy velocity and luminosity measurement techniques.  Furthermore, the choice of local comparison sample influences the interpretation of Tully-Fisher evolution.  Combined, these two effects change our Cl0024 TFR results by $\sim$1 mag.
\item We have probed the dependence of Cl0024 members' Tully-Fisher residuals upon other galaxy properties, in the hopes of uncovering evidence of the evolutionary processes that may be affecting the galaxies.  The strongest relationship we find is between TF residuals and rotation velocities in the cluster, suggesting Tully-Fisher slope evolution.  We caution that magnitude selection and the choice of Tully-Fisher slope used in the residual analysis may strongly affect this result, however.
\item Evidence for galaxy interactions as an evolutionary driver in Cl0024 is unclear from the weak correlations we find between TF residuals and photometric residuals and asymmetries.  No apparent relationship exists between TF residuals and emission line residuals and asymmetries, but this may be due to the relatively low resolution of our kinematic data.  Furthermore, this result may be affected by small number statistics.
\item We find no evidence for counter-rotation between stars (as measured from absorption) and gas (as measured from emission) in two Cl0024 members for which we have high enough S/N data to make such measurements.  One of these galaxies exhibits characteristics of an AGN; the other appears to be morphologically disturbed.  
\end{enumerate}

This study is intended as a pilot investigation of the Tully-Fisher relation in distant clusters.  With our small sample size and focus on a single cluster, we cannot make sweeping statements about the TFR and TF residuals as indicators of evolutionary processes in clusters.  However, our results are intended as a baseline for comparison to future studies made with larger samples of distant cluster galaxies.  

In the future, observations with integral field units (e.g., Pracy et al.\ 2005) will likely be essential to understanding the interplay between galaxy internal kinematics and evolutionary drivers in both the cluster and field environments.  Such observations will provide two-dimensional measurements linking star formation rates, metallicities, shock diagnostics, and kinematics, allowing observers to determine direct links between these diagnostics and evolutionary predictions.

\acknowledgments
\noindent We thank the anonymous referee for insightful comments which have greatly improved our analysis.  We are grateful to Sheila Kannappan, with whom we discussed the galaxy velocity measurement techniques that are used in local Tully-Fisher studies.  St\'{e}phane Courteau kindly provided us with local galaxy velocity measurements.  It is also a pleasure to thank Greg Wirth for help with the Cl0024 spectra, and Dan Kelson for distributing his Python spectral reduction modules and training us to use them.  We thank Sandy Faber for her input on our statistical analysis, and Kim-Vy Tran for helpful discussions on spectral reduction and GIM2D modeling.  Christopher Willmer provided restframe magnitudes and color estimates.  Archival HST images were retrieved from the Canadian Astronomy Data Centre, which is operated by the Herzberg Institute of Astrophysics, National Research Council of Canada.  Our data reduction procedure made use of the IRAF software package, distributed by the National Optical Astronomy Observatories, which are operated by the Association of Universities for Research in Astronomy, Inc., under cooperative agreement with the National Science Foundation.  We thank the Keck staff for their help in acquiring the spectra, and the Hawaiian people for the use of their sacred mountain.  Work on this project was supported by the following grants: HST STScI 09206.01-A and HST STScI 09555.01-A.  AJM appreciates support from the the National Science Foundation from grant AST-0302153 through the NSF Astronomy and Astrophysics Postdoctoral Fellows program.

\begin{figure}
\figurenum{1}
\epsscale{0.8}
\plotone{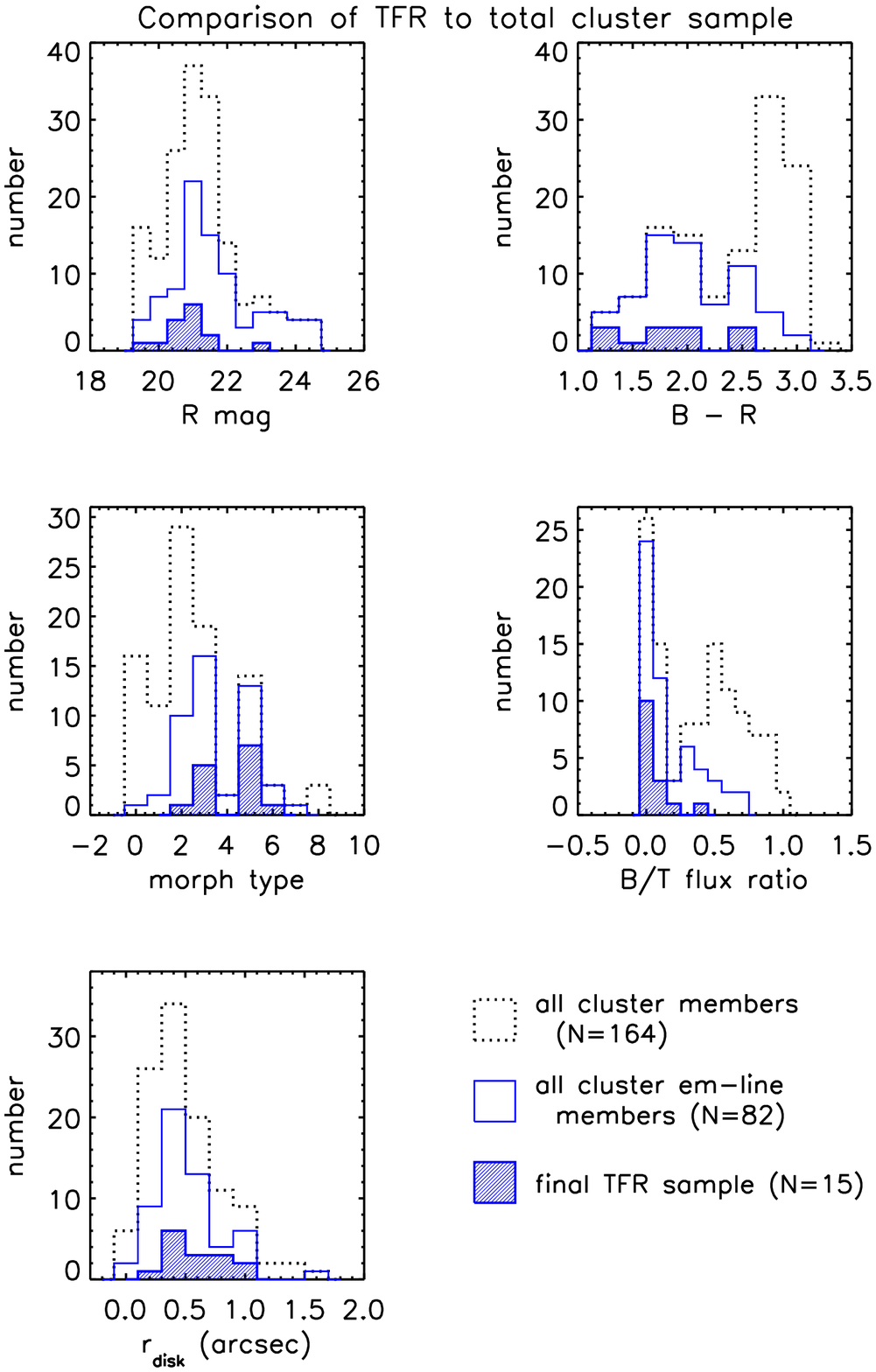}
\caption{Histograms of the distributions of galaxy $R$ magnitudes, $B-R$ colors, morphological types, bulge-to-total flux ratios (B/T), and disk scale lengths ($r_{d}$) for all cluster members in our Keck survey ({\it dotted}), all cluster emission line galaxies in our survey ({\it open}), and our final Tully-Fisher sample ({\it hatched}).  In general, galaxies in the final Tully-Fisher sample are relatively bright, blue, late-types with large disks.  Morphological types were obtained from Treu et al.\ 2003: 0 = E, 1 = E/S0, 2 = S0, 3 = Sa+b, 4 = S, 5 = Sc+d, 6 = Irr, 7 = Unclass, 8 = Merger.  Note that no $B - R$ colors were measured for TFR 01 or TFR 03; no morphological classification has been given to TFR 03.}
\end{figure}

\clearpage

\begin{figure}
\figurenum{2}
\epsscale{0.8}
\plotone{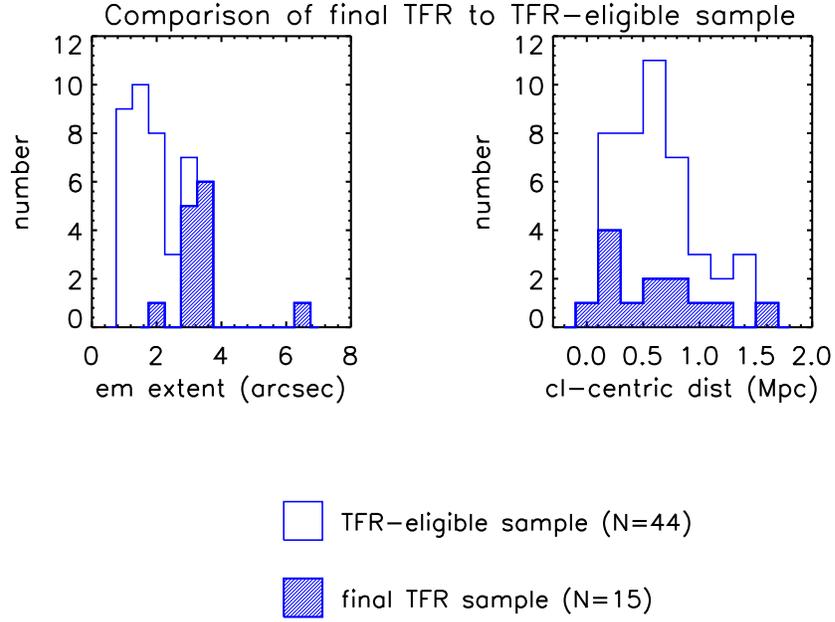}
\caption{Histograms of the distributions of galaxy emission line extents (full-width) and projected cluster-centric distances for our Tully-Fisher-eligible sample ({\it open}) and our final Tully-Fisher sample ({\it hatched}).  Tully-Fisher-eligible galaxies are emission line cluster members that lie within the HST-imaged region of the cluster and satisty our inclination requirement ($30^{\circ} < i < 80^{\circ}$) for rotation curve analysis.  Only 15 of the 44 eligible galaxies made it into our final Tully-Fisher sample.  In general, the galaxies in this final sample have the largest emission line extents, but both samples are similarly distributed within the cluster.}
\end{figure}

\clearpage

\begin{figure}
\figurenum{3}
\epsscale{0.7}
\begin{center}
\includegraphics[angle=90,width=5.0in]{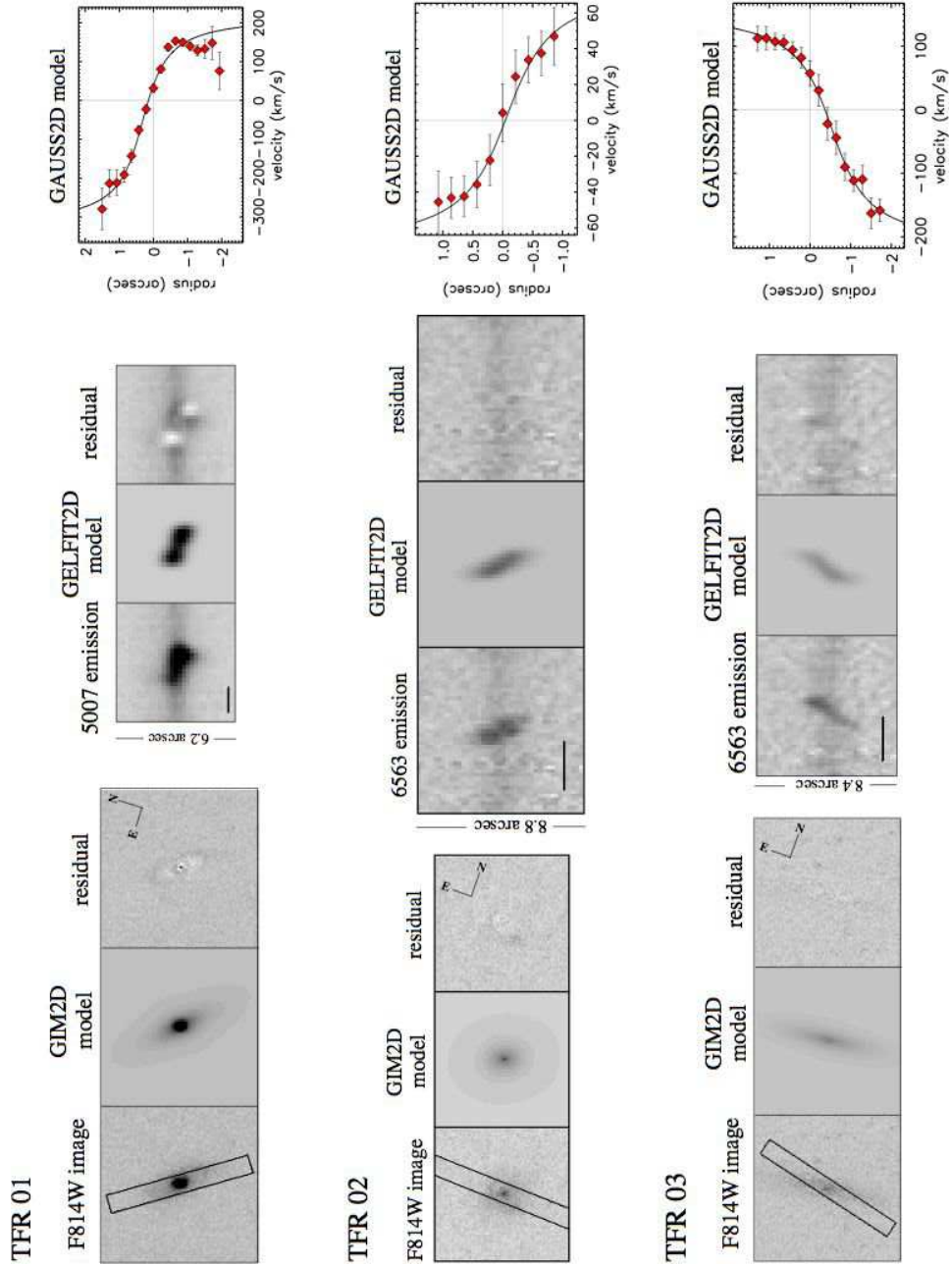} 
\caption{Data and rotation curve measurements. {\it Left:} WFPC2 F814W image of Cl0024 cluster member with slit position overlaid, GIM2D structural model of galaxy, residual (image -- model) image with compass rose indicating North and East.  F450W images have been included where available. {\it Middle:} 2D emission with continuum from Keck (LRIS) spectrum, GELFIT2D best-fit model of emission, residual image. {\it Right:} Data points are velocity measurements from Gaussian fits to emission; solid curve represents same analysis on best-fit GAUSS2D emission line model.  Within a given panel, images are shown at the same contrast level.  For a given object, panels are shown to scale.  Black horizontal line in lower left corner of 2D emission depicts a velocity scale of 500 km s$^{-1}$.}
\end{center}
\end{figure}

\begin{figure}
\figurenum{3}
\epsscale{0.7}
\begin{center}
\includegraphics[angle=90,width=5.0in]{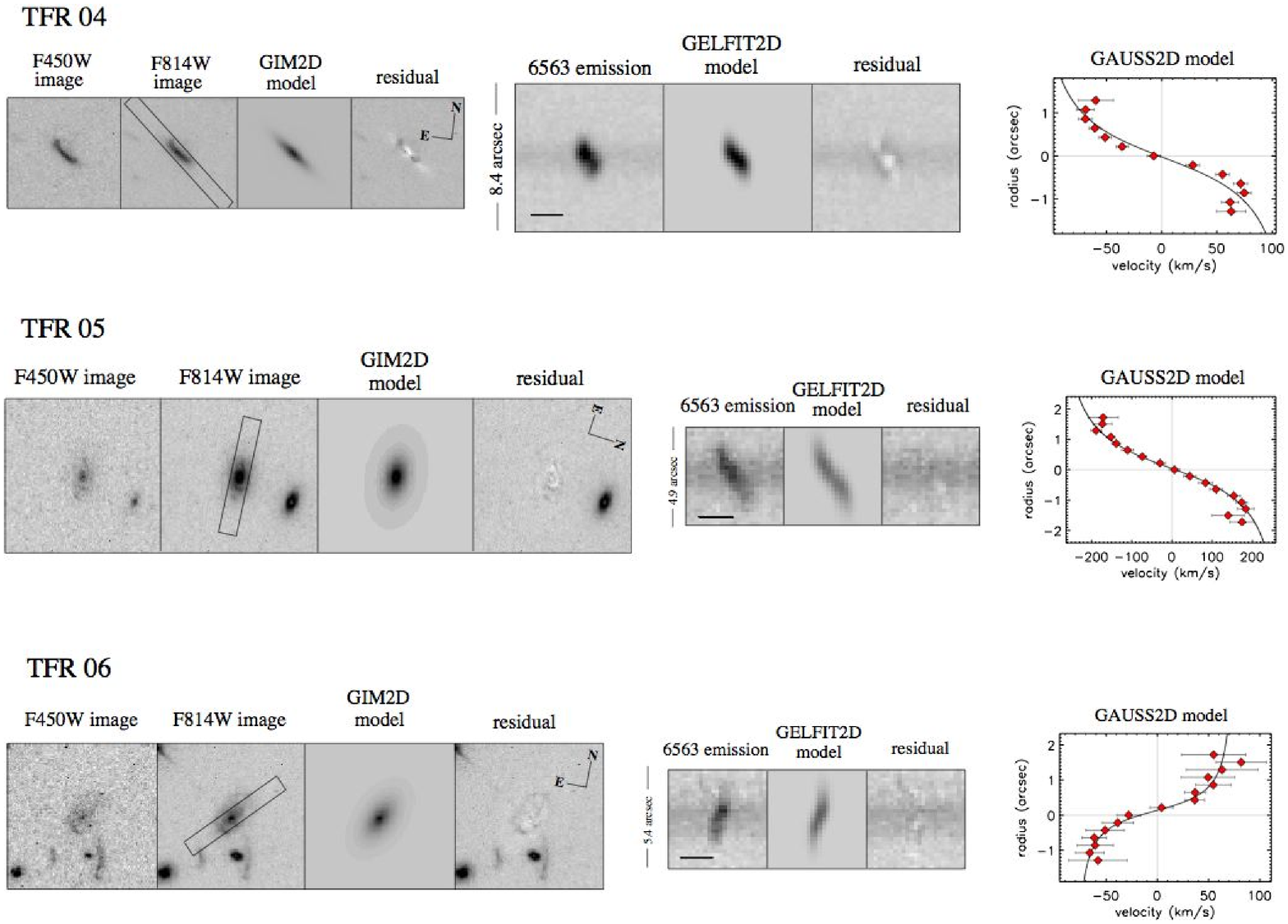} 
\caption{Rotation curve measurements, continued.}
\end{center}
\end{figure}

\begin{figure}
\figurenum{3}
\epsscale{0.7}
\begin{center}
\includegraphics[angle=90,width=5.0in]{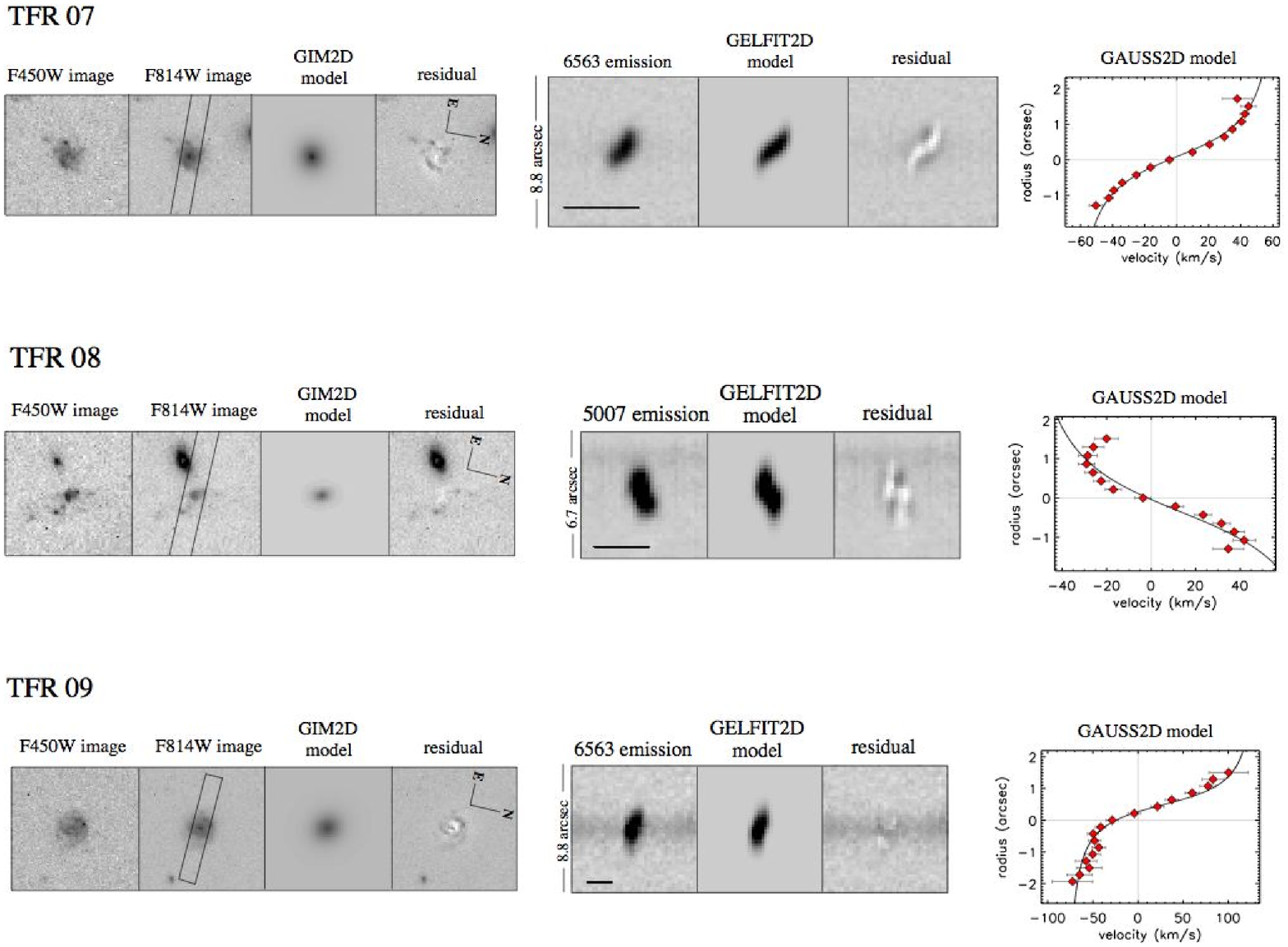} 
\caption{Rotation curve measurements, continued.}
\end{center}
\end{figure}

\begin{figure}
\figurenum{3}
\epsscale{0.7}
\begin{center}
\includegraphics[angle=90,width=5.0in]{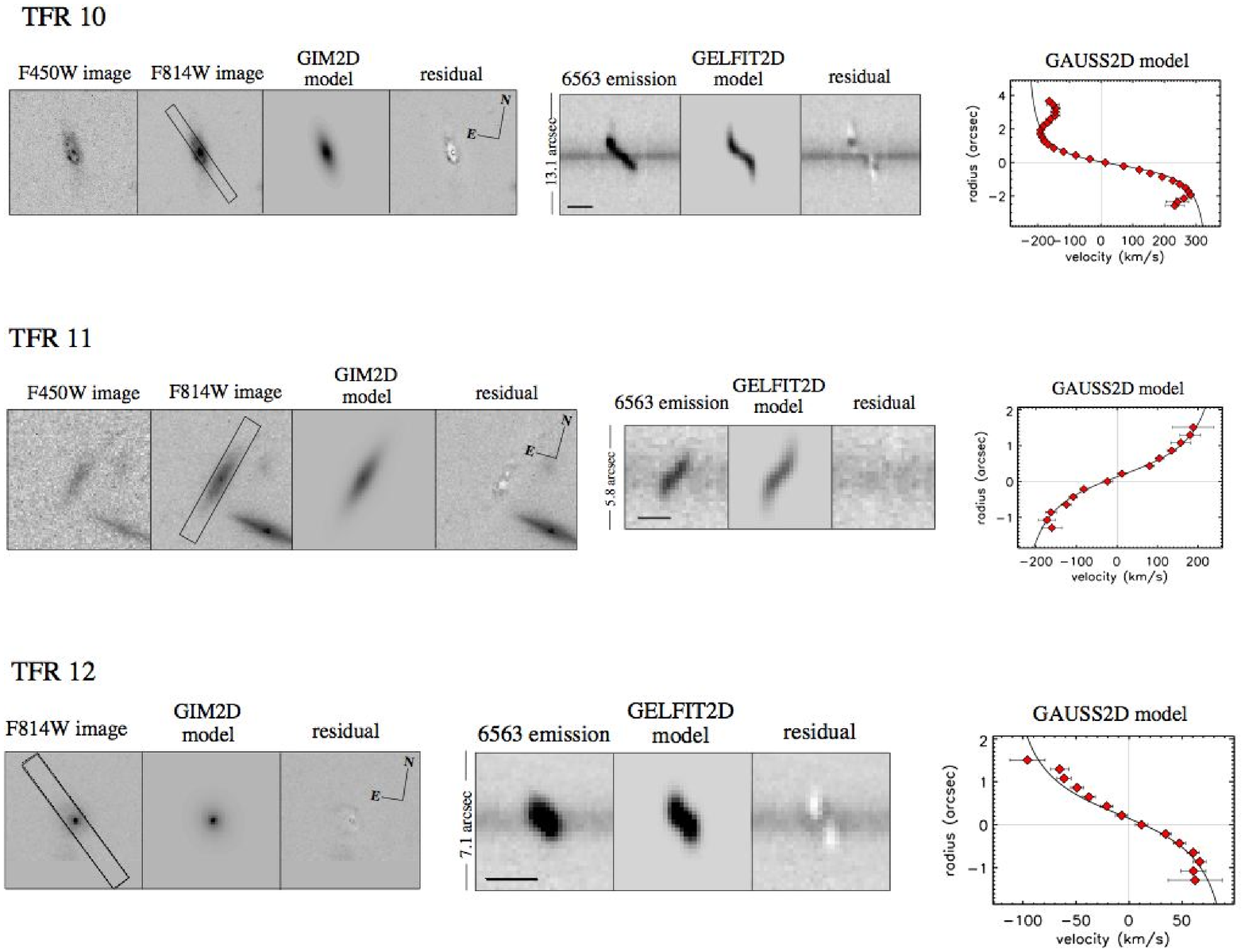} 
\caption{Rotation curve measurements, continued.}
\end{center}
\end{figure}

\begin{figure}
\figurenum{3}
\epsscale{0.7}
\begin{center}
\includegraphics[angle=90,width=5.0in]{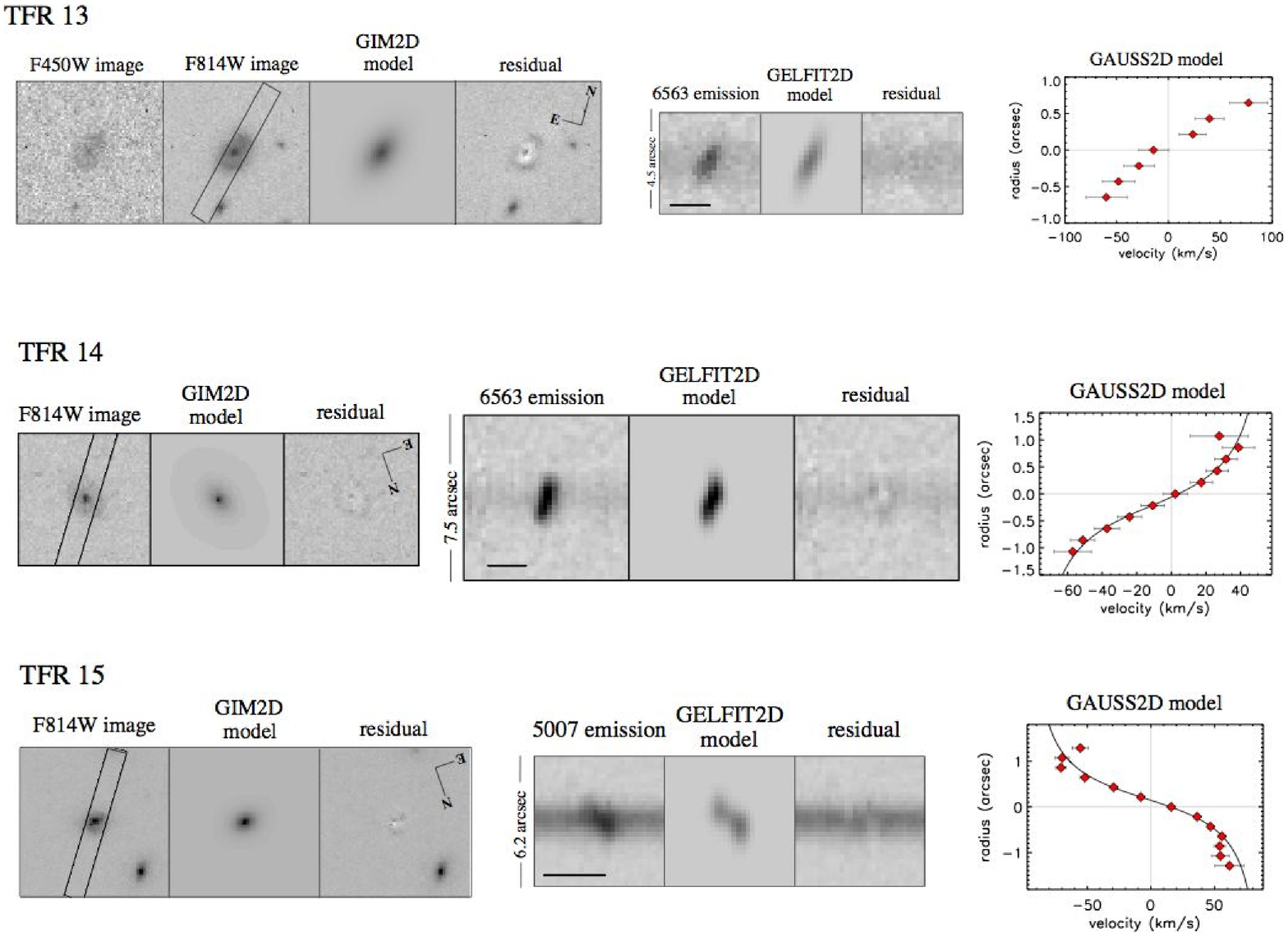} 
\caption{Rotation curve measurements, continued.}
\end{center}
\end{figure}

\clearpage

\begin{figure}
\figurenum{4}
\epsscale{0.8}
\plotone{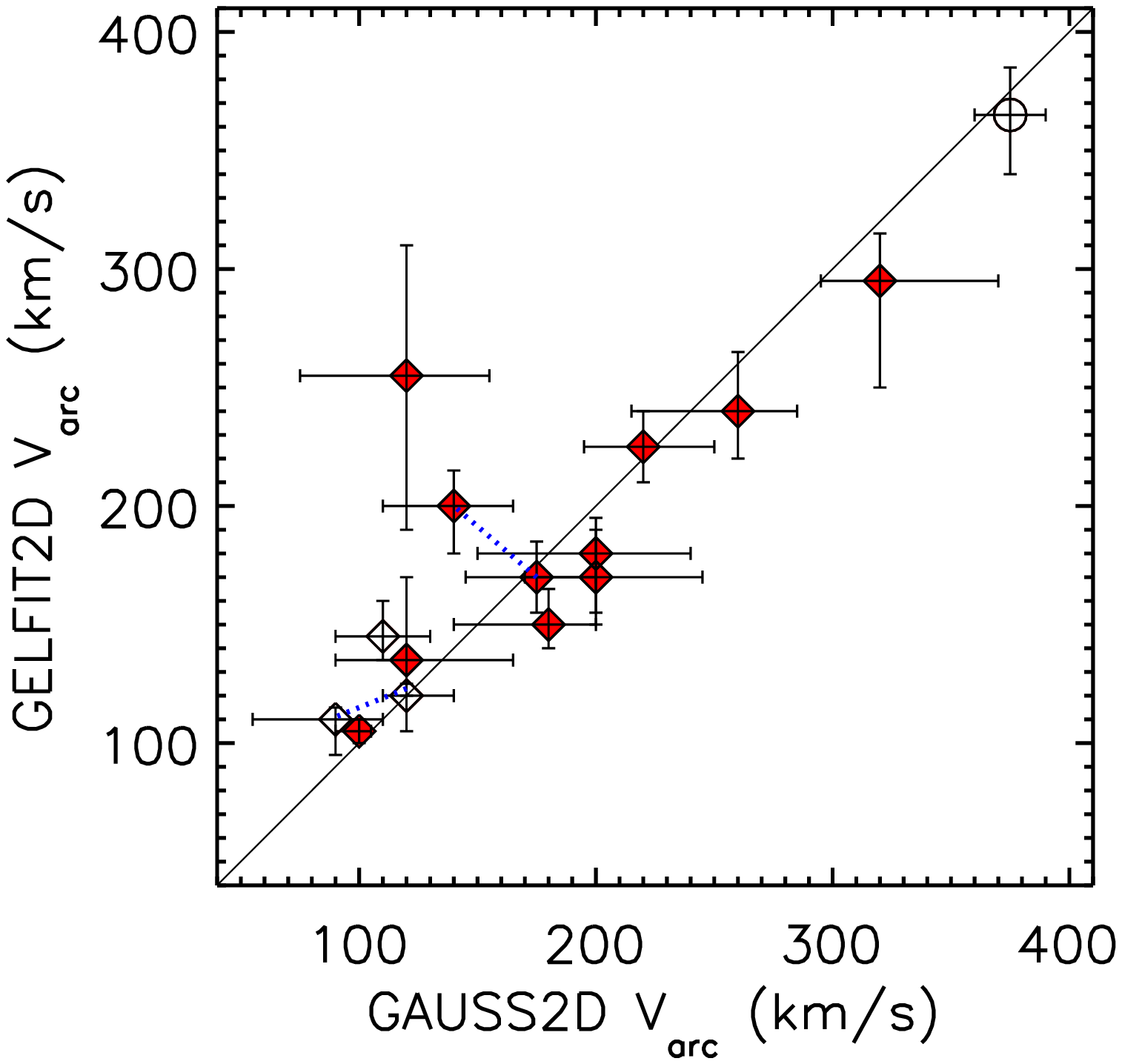}
\caption{Comparison of GAUSS2D and GELFIT2D velocity measurements for our sample of 15 Cl0024 members demonstrates good agreement between the two analysis techniques.  Duplicate measurements are shown, linked by a dotted lines, for the two galaxies (TFR 04 and TFR 15) for which we had two exposures at different slit PAs.  Two galaxies (TFR 01 and TFR 13) are not represented due a poor fit from one of models.  Solid symbols represent ``normal" sample.  Open diamonds represent ``distorted" galaxies (TFR 04 and TFR 08).  Open circle denotes TFR 10, which exhibits a knotty ring of emission.  Solid line represents $V_{arc}^{GAUSS2D} = V_{arc}^{GELFIT2D}$.}
\end{figure}

\clearpage

\begin{figure}
\figurenum{5}
\epsscale{0.8}
\plotone{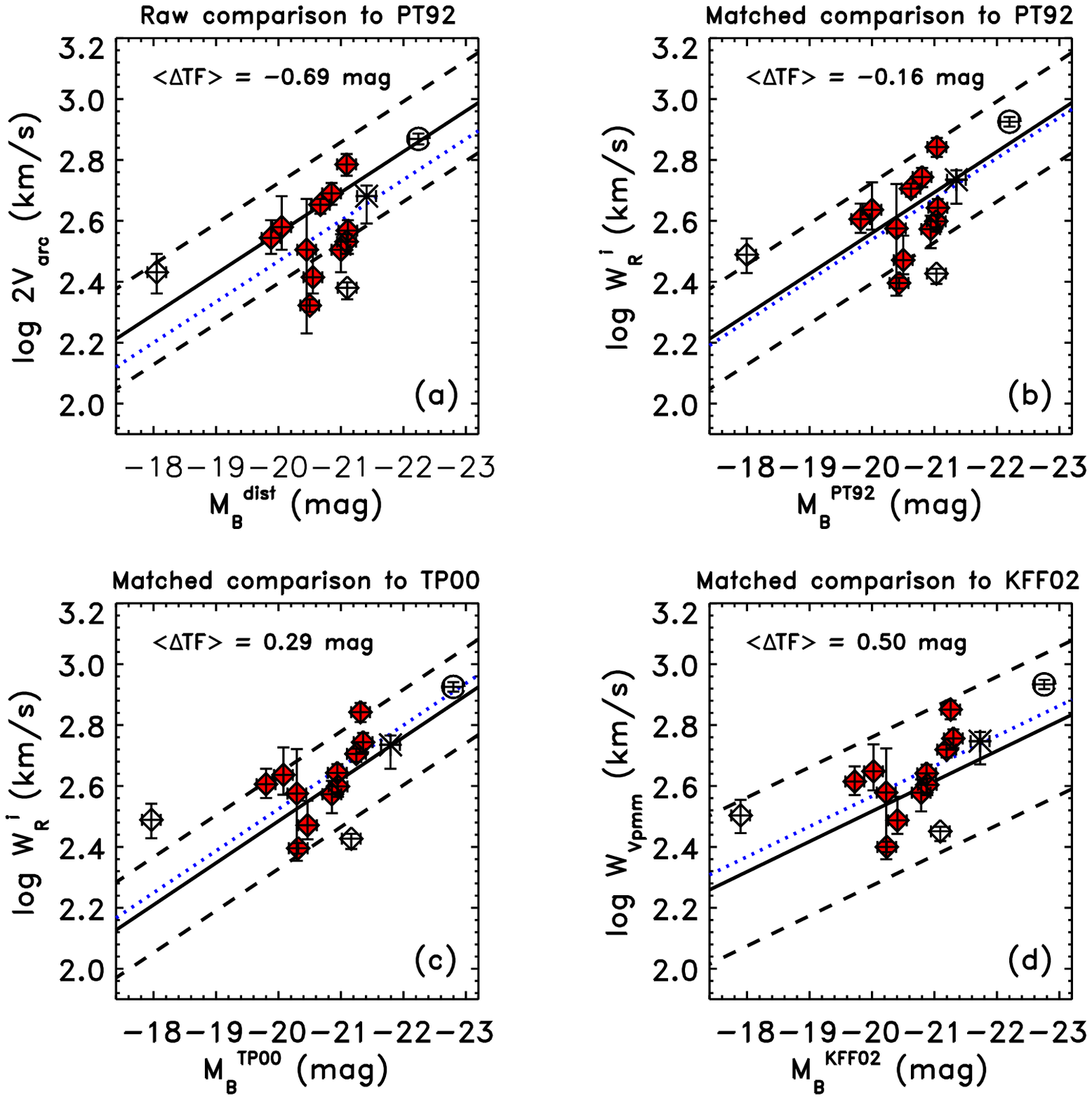}
\caption{Comparisons of the Cl0024 TFR to the TFR derived for local field galaxy samples.  {\it Panel (a)}: a ``raw" comparison of the Cl0024 TFR to the PT92 local relation, without accounting for differences between velocity and luminosity measurement techniques (see text). {\it Panel (b)}: a ``matched" comparison to PT92, in which systematic differences between measurement techniques have been accounted for. {\it Panels (c) and (d)}: ``matched" comparisons to the local TP00 and KFF02 samples.  In each panel, solid symbols represent ``normal" Cl0024 members; open diamonds denote ``distorted'' galaxies (TFR 04 and TFR 08); an open circle denotes TFR 10, which exhibits a knotty ring of emission; and a star represents TFR 01, a likely AGN.  Solid and dashed lines show the local relations with 3$\sigma$ errors.  The dotted line demonstrates the best-fit biweight offset between our ``normal'' data points and the local relations (assuming local slope).  The magnitude of these offsets (in luminosity) is listed at the top of each panel.  From these plots, it is clear that the choice of comparison sample and the method of comparison (``raw" or ``matched") have an important effect on the interpretation of Tully-Fisher evolution.}
\end{figure}

\clearpage

\begin{figure}
\figurenum{6}
\epsscale{0.6}
\plotone{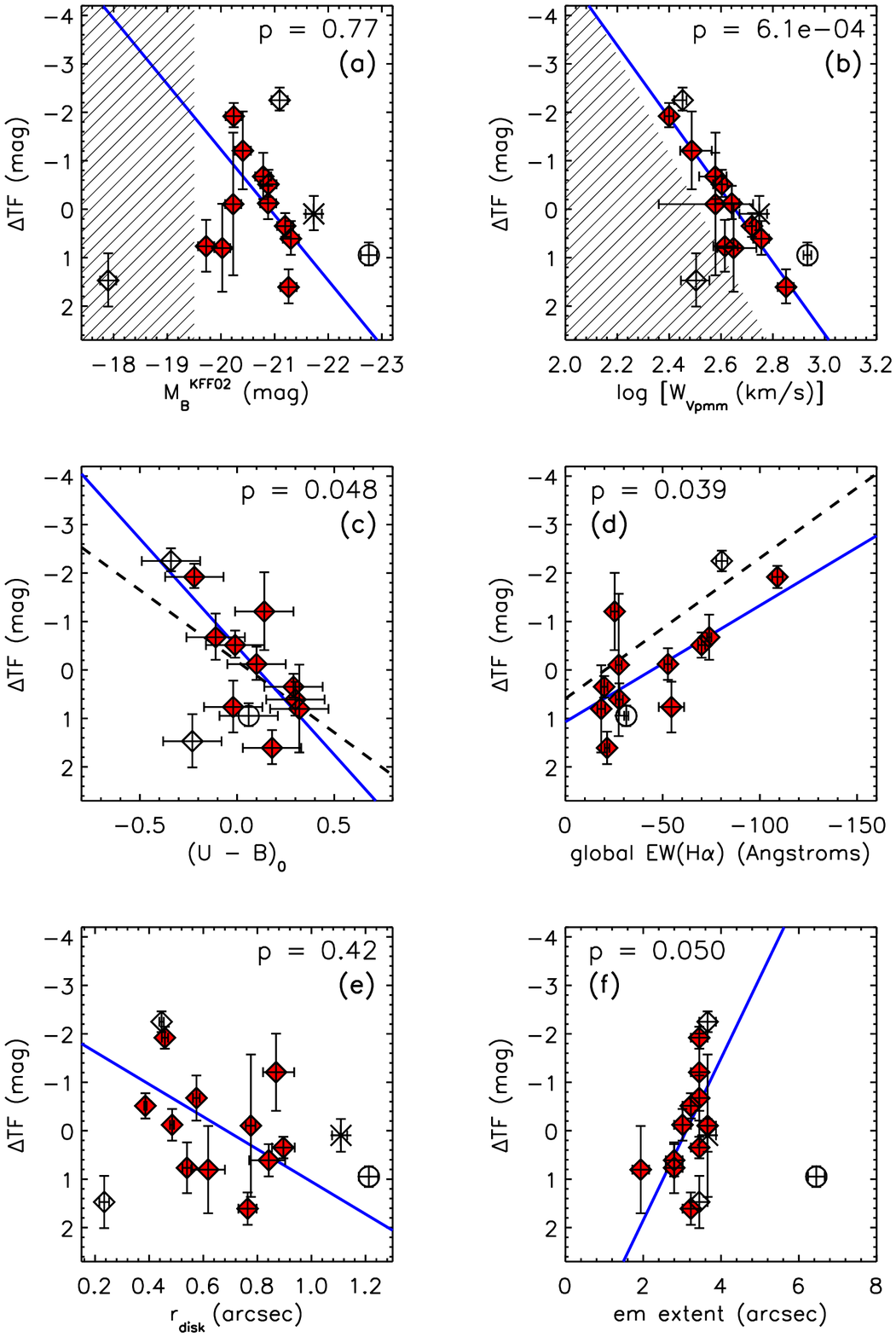}
\caption{Cl0024 residuals from the TFR (a simple 0.50 mag offset from the local KFF02 Tully-Fisher relation) are shown against luminosity ($M_{B}^{\rm KFF02}$), rotation velocity (log $W_{V_{\rm pmm}}$), $(U-B)_{0}$ colors, global EW(H$\alpha$), $r_{disk}$, and emission line extent (panels {\it a} through {\it f}, respectively).  Panels ({\it a}) and ({\it b}) are shaded to indicate that our sample is incomplete at $M_{B} > -19.5$.  No relation between TF residuals and luminosity is found; however, Cl0024 TF residuals appear to be correlated with galaxy rotation velocities.  Loose correlations with $(U-B)_{0}$ and $r_{disk}$ are seen.  Stronger correlations exist with EW(H$\alpha$) and emission line extents.  In all panels, solid data points represent the ``normal" sample; open diamonds are ``distorted" galaxies (TFR 04 and TFR 08); an open circle denotes TFR 10, which exhibits a ring of emission; a star denotes TFR 01, a likely AGN.  Solid lines represent fits to the ``normal" sample only.  The chance probabilities of these correlations, determined using the Spearman rank test, are noted in the upper right corner of each panel.  Correlations are only considered significant if p $<$ 0.05, indicating 95\% confidence the correlation did not occur by chance.  In panel ({\it c}), a dashed line shows the relationship between $\Delta$TF and $(U-B)_{0}$ found by KFF02 using a local galaxy sample.  In panel ({\it d}), the dashed line shows the relationship between $\Delta$TF and EW(H$\alpha$) found by KFF02.  Note that no colors were measured for TFR 01 or TFR 02; EW(H$\alpha$) was not measured for TFR01 or TFR08.}
\end{figure}

\clearpage

\begin{figure}
\figurenum{7}
\epsscale{0.9}
\plotone{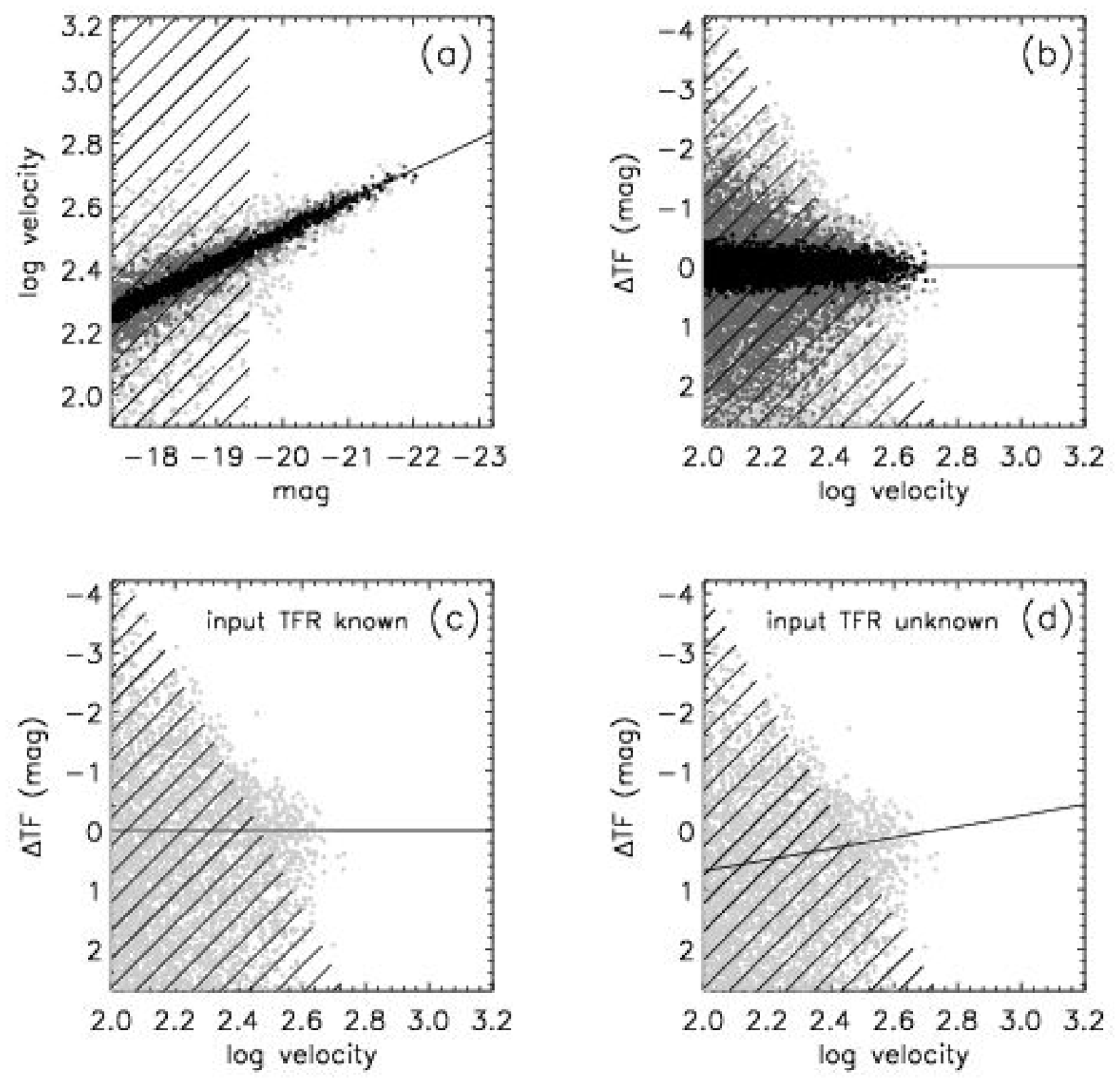}
\caption{ Panel {\it (a):} Tully-Fisher relation for three randomly-generated samples: low-dispersion (black points), medium-dispersion (dark grey points), and high-dispersion (light grey points).  Solid black line depicts the $B$-band Tully-Fisher relation measured by KFF02; this is the input TFR for each of the generated samples.  Shading on the left side of the diagram indicates our magnitude bias against galaxies with $M_{B} > -19.5$.  For a given low velocity, only galaxies with relatively high luminosities make it into the unshaded area of the diagram, possibly biasing measurements of the Tully-Fisher slope.  Panel {\it (b):} The relationship between velocity and Tully-Fisher residuals ($\Delta$TF) is shown for the three randomly generated samples.  This diagram is also shaded to reflect a magnitude bias.  The solid line at $\Delta$TF = 0 represents the fact that no Tully-Fisher offset was input into the samples before dispersion was added.  Some low-velocity, high-luminosity galaxies make it into the upper left of the unshaded region of the diagram.  Panels {\it (c)} and {\it (d)} again show velocity versus $\Delta$TF for the high dispersion sample.  In panel {\it (c)}, $\Delta$TF is determined from the known input TFR, whereas in panel {\it (d)}, $\Delta$TF is determined against the {\it measured} Tully-Fisher relation.  In both panels, the shaded region indicates incompleteness at $M_{B} > -19.5$, and a solid line gives the true velocity-$\Delta$TF distribution of the sample. }
\end{figure}

\clearpage
\begin{figure}
\figurenum{8}
\epsscale{0.6}
\plotone{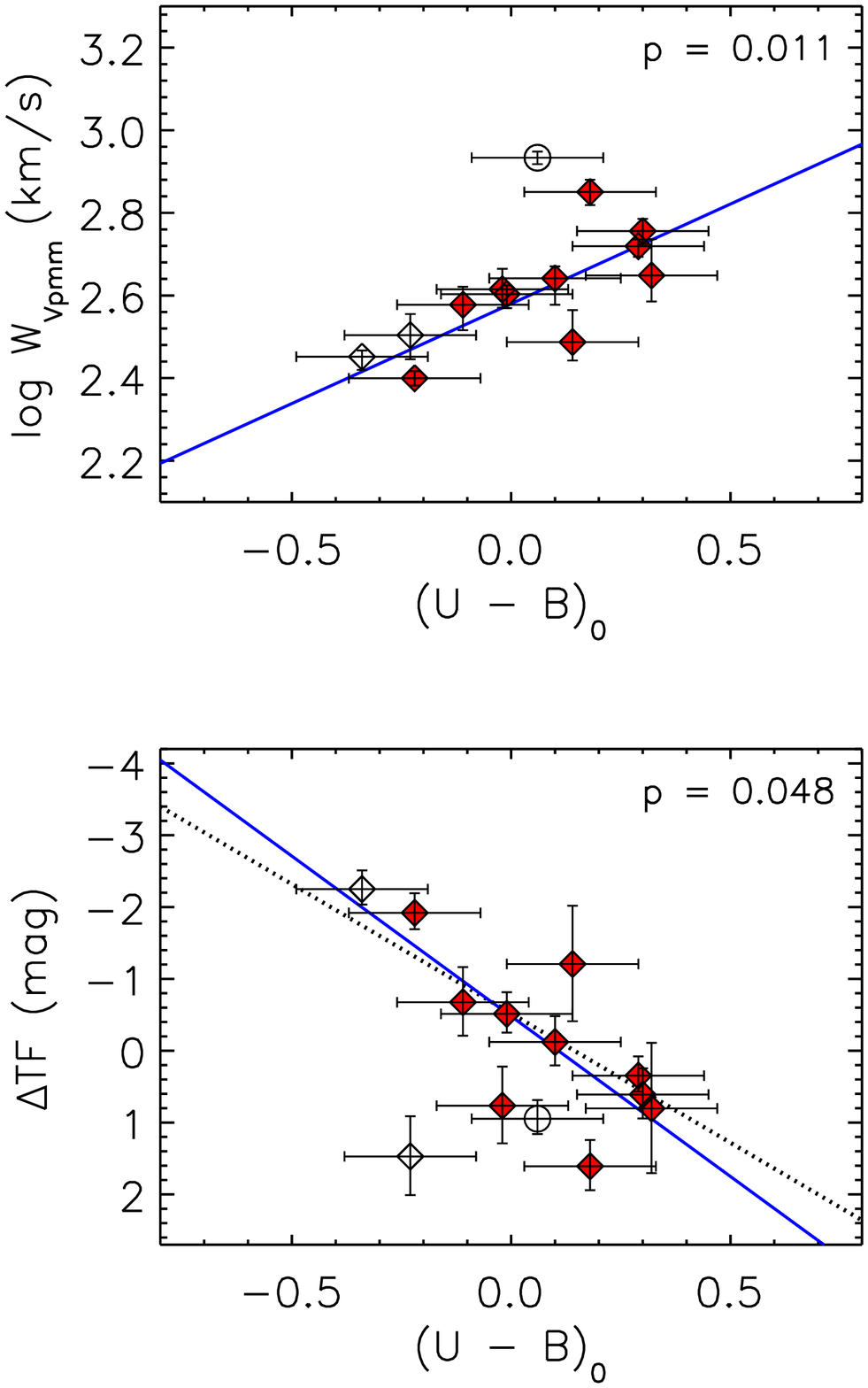}
\caption{ {\it Top panel:} $(U-B)_{0}$ colors and rotation velocities ($W_{V_{\rm pmm}}$) are correlated for the Cl0024 TFR sample, demonstrating that it is difficult to decouple the relationships between the velocities, luminosities, and colors of these objects. {\it Bottom panel:} We again plot Cl0024 Tully-Fisher residuals against $(U-B)_{0}$ colors.  The dotted line in this plot represents the color-$\Delta$TF relation derived by comparing the velocity-$\Delta$TF and color-velocity relations.  The color-$\Delta$TF relation derived in this way is close to our line fit to the data points (solid line).  This indicates that the color-$\Delta$TF relation in Cl0024 may simply be a natural outcome of the velocity-$\Delta$TF and color-velocity relations.  In both plots, solid data points represent the ``normal" sample; open diamonds are ``distorted" galaxies (TFR 04 and TFR 08); an open circle denotes TFR 10, which exhibits a ring of emission.  Solid lines represent fits to the ``normal" sample only.  Chance probabilities of these correlations, determined using the Spearman rank test, are noted in the upper right corner of each panel.  Note that no colors were measured for TFR 01 or TFR 02.}
\end{figure}

\clearpage

\begin{figure}
\figurenum{9}
\epsscale{0.8}
\plotone{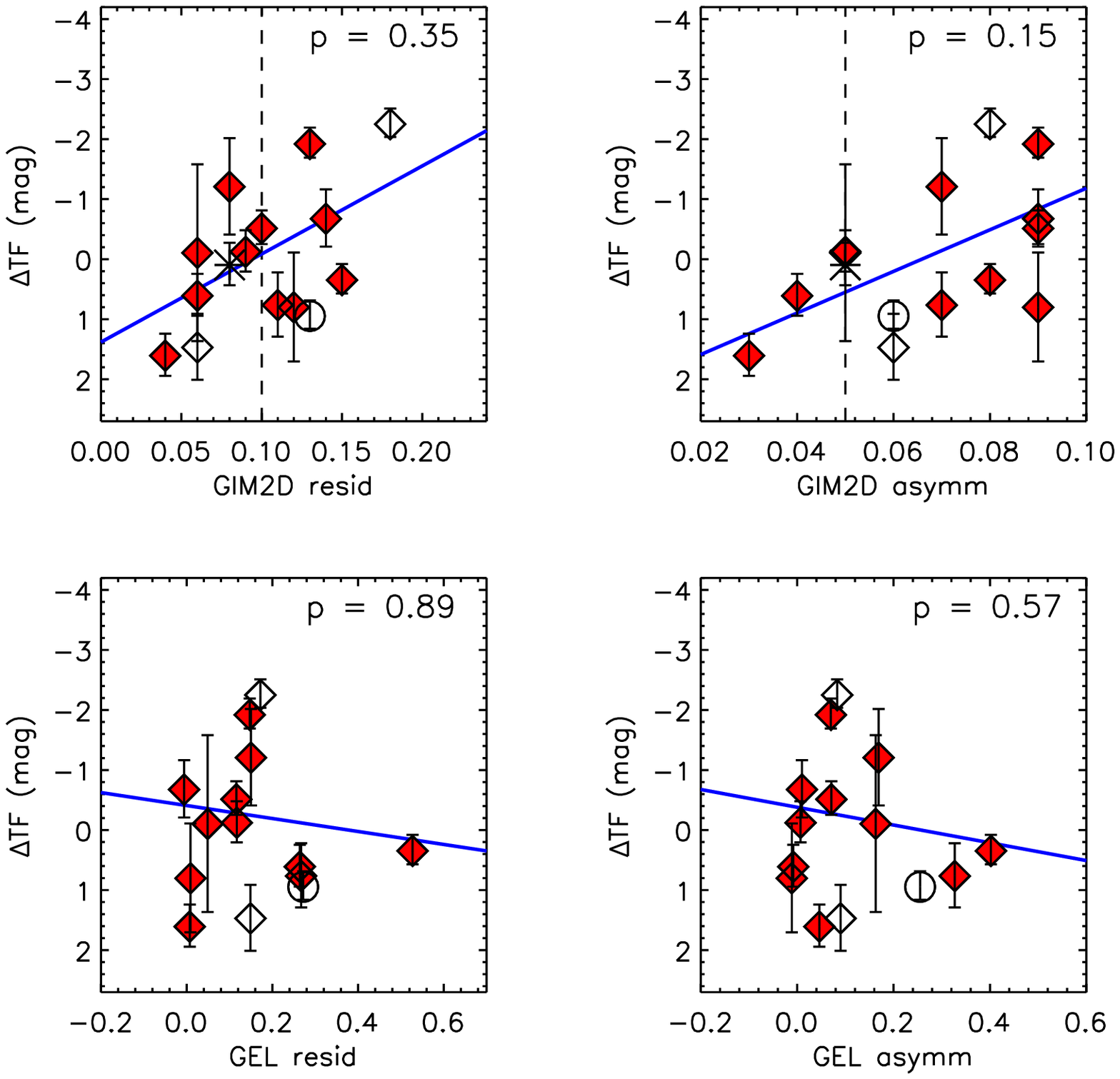}
\caption{Cl0024 residuals from the local Tully-Fisher relation are shown against residual and asymmetry parameters from GIM2D and GELFIT2D.  Very loose correlations between TF residuals and GIM2D residuals and asymmetries can be seen ({\it top panels}) such that galaxies with high residuals and asymmetries tend to be overluminous.  Dotted vertical lines in these panels show the cutoff between low and high residuals and asymmetries as defined by Tran et al. (2001, 2003).  Line fits between TF residuals and GELFIT2D residuals and asymmetries are poor and are consistent with no correlation.  Solid data points represent the ``normal" sample; open diamonds are ``distorted" galaxies (TFR 04 and TFR 08); an open circle denotes TFR 10, which exhibits a ring of emission; a star denotes TFR 01, a likely AGN.  Solid lines represent fits to the ``normal" sample only.   Chance probabilities of these correlations, determined using the Spearman rank test, are noted in the upper right corner of each panel.  Note that no GELFIT2D information is included for TFR 01, which was poorly fit using this technique.}
\end{figure}

\clearpage

\begin{figure}
\figurenum{10}
\epsscale{0.8}
\plotone{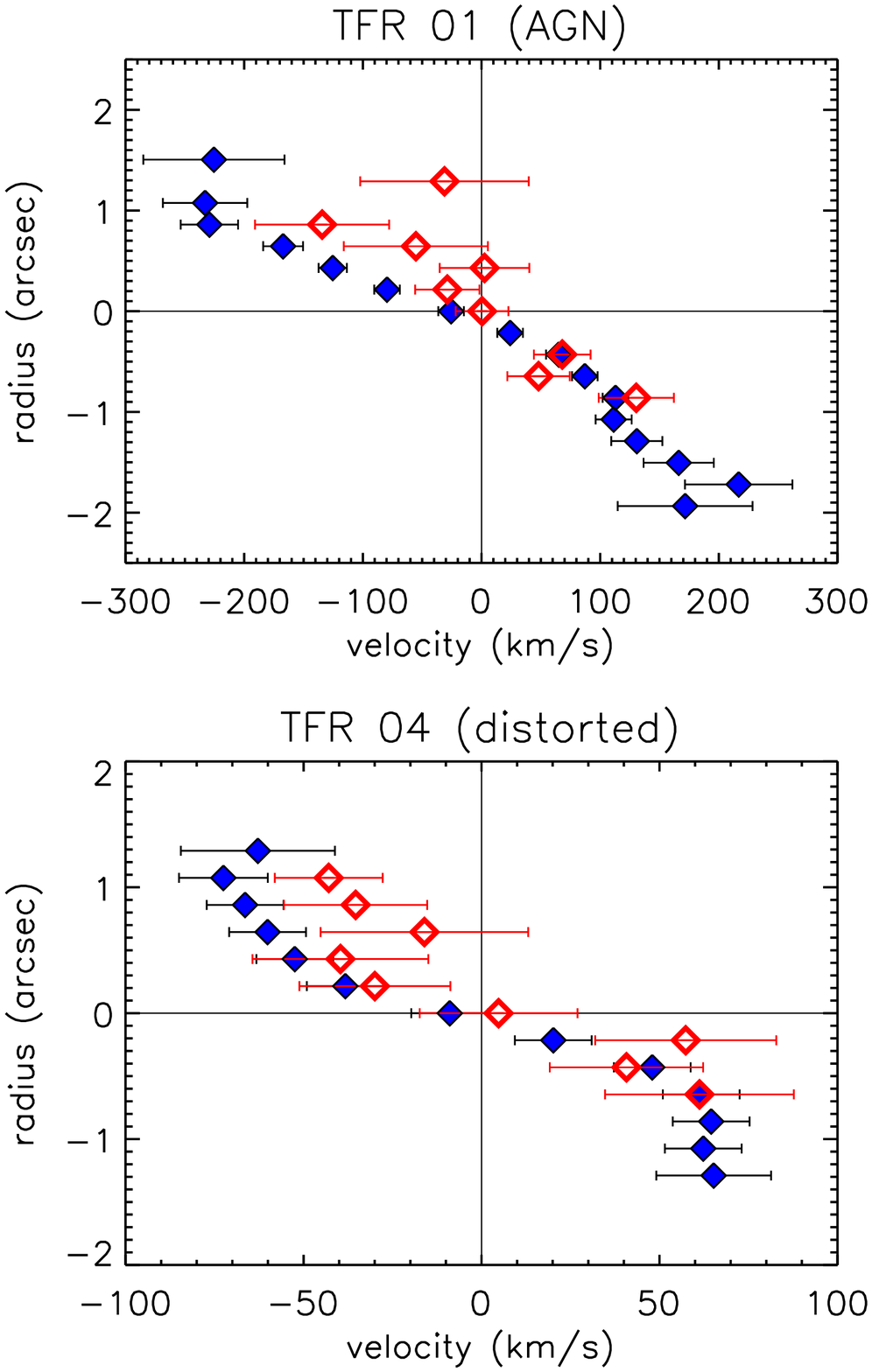}
\caption{Raw rotation velocity measurements from Gaussian fits to emission and absorption lines are shown for two Cl0024 members: TFR 01, a likely AGN, and TFR 04, a ``distorted" galaxy.  Open data points represent measurements from absorption; solid data points represent measurements from emission.  No clear evidence for counter-rotation between stars (absorption) and gas (emission) is seen for either galaxy.}
\end{figure}

\clearpage

\begin{deluxetable}{llrcrr} 
\tablecolumns{6} 
\tablewidth{0pc} 
\tablecaption{Comparison of TFR and Cluster Sample Properties} 

\tablehead{\colhead{Property} & \colhead{Sample} & \colhead{Mean} & \colhead{Std.\ Dev.} & \colhead{Min} & \colhead{Max}}

\startdata
$R$ mag & all cluster & 21.27 & 1.16 & 19.31 & 24.35 \\
 & cluster em & 21.64 & 1.27 & 19.65 & 24.35 \\
 & TFR sample & 20.88 & 0.80 & 19.69 & 23.20 \\
\tableline
$B - R$ & all cluster & 2.40 & 0.54 & 1.14 & 3.21 \\
 & cluster em & 2.06 & 0.46 & 1.14 & 2.94 \\
 & TFR sample & 1.84 & 0.45 & 1.20 & 2.47 \\ 
\tableline
Morph\tablenotemark{a} & all cluster & 2.58 & 1.97 & 0.00 & 8.00 \\
 & cluster em & 3.50 & 1.54 & 0.00 & 7.00 \\
 & TFR sample & 4.14 & 1.23 & 2.00 & 6.00 \\  
\tableline
B/T & all cluster & 0.38 & 0.31 & 0.00 & 1.00 \\
 & cluster em & 0.17 & 0.20 & 0.00 & 0.70 \\
 & TFR sample & 0.07 & 0.10 & 0.00 & 0.40 \\ 
\tableline
$r_{d}$ (\arcsec) & all cluster & 0.49 & 0.32 & 0.01 & 1.66 \\
 & cluster em & 0.50 & 0.29 & 0.01 & 1.66 \\
 & TFR sample & 0.57 & 0.22 & 0.25 & 0.99 \\ 
\enddata
\tablenotetext{a}{Morphological types from Treu et al.\ 2003: 0 = E, 1 = E/S0, 2 = S0, 3 = Sa+b, 4 = S, 5 = Sc+d, 6 = Irr, 7 = Unclass, 8 = Merger}
\end{deluxetable}



\begin{deluxetable}{llrcrr} 
\tablecolumns{6} 
\tablewidth{0pc} 
\tablecaption{Comparison of TFR and TF-Eligible Sample Properties} 

\tablehead{\colhead{Property} & \colhead{Sample} & \colhead{Mean} & \colhead{Std.\ Dev.} & \colhead{Min} & \colhead{Max}}

\startdata
em extent (\arcsec)& TF eligible & 2.16 & 1.10 & 0.86 & 6.45 \\
 & TFR sample & 3.44 & 1.02 & 1.94 & 6.45 \\
\tableline
cl-centric dist ($Mpc$) & TF eligible & 0.65 & 0.37 & 0.03 & 1.56 \\
 & TFR sample & 0.61 & 0.45 & 0.03 & 1.56 \\ 
\enddata
\end{deluxetable}

\clearpage

\begin{landscape}

\begin{deluxetable}{lrrrrrrrrrrrrr} 
\tablecolumns{14} 
\tablewidth{0pc} 
\tabletypesize{\scriptsize}
\tablecaption{Cl0024 Tully-Fisher Sample: Basic Data} 

\tablehead{\colhead{object} & 
\colhead{R.\ A.} & 
\colhead{Dec.} & 
\colhead{$z$} &
\colhead{Morph} & 
\colhead{B/T} &
\colhead{$r_{d}$} &
\colhead{$\Delta$PA} &
\colhead{{\it inc}} &
\colhead{$R^{obs}$} & 
\colhead{($B-R$)$^{obs}$} & 
\colhead{EW(H$\alpha$)} &
\colhead{em extent} & 
\colhead{D$_{cl}$}\\
\colhead{} & 
\colhead{(J2000)} & 
\colhead{(J2000)} & 
\colhead{} & 
\colhead{} & 
\colhead{} & 
\colhead{($\arcsec$)} & 
\colhead{($deg$)} & 
\colhead{($deg$)} & 
\colhead{($mag$)} & 
\colhead{($mag$)} &
\colhead{(\AA)} & 
\colhead{($\arcsec$)} & 
\colhead{($Mpc$)} \\
\colhead{(1)} & 
\colhead{(2)} & 
\colhead{(3)} & 
\colhead{(4)} & 
\colhead{(5)} & 
\colhead{(6)} & 
\colhead{(7)} & 
\colhead{(8)} & 
\colhead{(9)} & 
\colhead{(10)} & 
\colhead{(11)} & 
\colhead{(12)} & 
\colhead{(13)} & 
\colhead{(14)}  }  
\startdata
TFR 01	 & 0:26:19.41	 & +17:07:03.2	 & 0.3967	 & S0		& $0.40$	& 1.11	& --10		& 70	& 20.08	 & $\cdots$ & $\cdots$ & 3.7	& 1.56 \\
TFR 02	 & 0:26:31.03	 & +17:07:21.8	 & 0.3804	 & Sc+d		& $0.01$	& 0.78	& --29		& 37	& 20.72	 & $\cdots$	& $-27.4 \pm 1.5$ & 3.7  	& 0.84	 \\
TFR 03	 & 0:26:27.14	 & +17:08:23.8	 & 0.3801	 &  $\cdots$	& $0.02$	& 0.90	& --20		& 78	& 20.89	 & 2.47	 	& $-19.9 \pm 1.7$ & 3.4	 	& 0.80 \\
TFR 04	 & 0:26:33.85	 & +17:08:56.6	 & 0.3987	 & Sc+d		& $0.00$	& 0.45	& --13,--8	& 79	& 20.76  & 1.20		& $-80.4 \pm 2.9$ & 3.7		& 0.29 \\
TFR 05	 & 0:26:33.54	 & +17:09:24.4	 & 0.3917	 & Sa+b		& $0.08$	& 0.76	& +1		& 58	& 20.18	 & 2.12	     	& $-21.4 \pm 1.2$ & 3.2	 	& 0.20 \\
TFR 06	 & 0:26:39.85	 & +17:09:28.8	 & 0.3854	 & Sc+d		& $0.02$	& 0.87	& --26		& 54	& 20.72	 & 1.98	     	& $-25.3 \pm 2.0$ & 3.4	 	& 0.34 \\
TFR 07	 & 0:26:42.45	 & +17:09:32.2	 & 0.3992	 &  Sc+d		& $0.00$	& 0.46	& --20		& 42	& 20.79	 & 1.30	 & $-109.0 \pm 2.7$	& 3.4	 	& 0.54 \\
TFR 08	 & 0:26:35.86	 & +17:09:38.7	 & 0.3966	 & Irr		& $0.10$	& 0.23	& +60		& 52	& 23.20	 & 1.30\tablenotemark{a}	& $\cdots$ & 3.4 	& 0.03	 \\
TFR 09	 & 0:26:42.64	 & +17:09:51.1	 & 0.3931	 & Sc+d		& $0.00$	& 0.58	& +41		& 39	& 20.21	 & 1.49	 	& $-73.9 \pm 1.1$ & 3.4	 	& 0.56	 \\
TFR 10	 & 0:26:41.92	 & +17:09:53.2	 & 0.3936	 & Sa+b		& $0.16$	& 1.21	& +17		& 68	& 19.20	 & 1.97	 	& $-31.3 \pm 1.1$ & 6.5	 	& 0.50 \\
TFR 11	 & 0:26:39.04	 & +17:10:02.3	 & 0.3943	 & Sc+d		& $0.05$	& 0.82	& +2		& 74	& 20.70	 & 2.47	 	& $-27.7 \pm 1.8$ & 2.8	 	& 0.29	  \\
TFR 12	 & 0:26:22.08	 & +17:10:06.0	 & 0.3963	 & Sa+b		& $0.07$	& 0.49	& +16		& 32	& 20.06	 & 1.87		& $-52.8 \pm 1.9$ & 3.0	 	& 1.10 \\
TFR 13	 & 0:26:36.99	 & +17:10:12.5	 & 0.3957	 & Sc+d		& $0.01$	& 0.62	& +2		& 55	& 21.20	 & 2.38	 	& $-18.4 \pm 2.0$ & 1.9	 	& 0.19 \\
TFR 14	 & 0:26:28.83	 & +17:10:50.3	 & 0.3910	 & Sa+b		& $0.04$	& 0.54	& --55		& 45	& 21.32	 & 1.66	 	& $-54.5 \pm 6.5$ & 2.8	 	& 0.66	  \\
TFR 15	 & 0:26:25.74	 & +17:12:05.8	 & 0.3931	 & Sa+b		& $0.03$	& 0.39	& +43,+52	& 39	& 20.11	 & 1.68	 	& $-70.0 \pm 1.8$ & 3.2	 	& 1.10	 \\
\enddata
\tablecomments{Columns: (1) object identification number (see Appendix for corresponding IDs from previous studies); (2) Right Ascension; (3) Declination; (4) redshift; (5) visual morphological classification from Treu et al.\ 2003; (6) $I_{814}$ bulge-to-total flux ratio; (7) $B_{450}$ disk scale length (or 1.2r$_{d}$ from $I_{814}$ in cases where $B_{450}$ imaging was not available); (8) offset of slit position angle from galaxy major axis, where a positive value means the offset is toward the East; (9) disk inclination from $I_{814}$ image; (10) observed total $R$-band magnitude, corrected for Galactic extinction; (11) observed $B-R$ color determined within a 3\arcsec\ aperture, corrected for Galactic extinction; (12) global H$\alpha$ equivalent width; (13) full-width emission line extent; (14) projected cluster-centric distance.}
\tablenotetext{a}{Roughly estimated from the $g - r$ color given in Koo et al.\ 1997, transformed to $B - R$ using colors given in Fukugita, Shimasaku, \& Ichikawa 1995. }
\end{deluxetable}

\clearpage
\end{landscape}

\begin{deluxetable}{rlll} 
\tablecolumns{4} 
\tablewidth{0pc} 
\tabletypesize{\scriptsize}
\tablecaption{Rotation Velocity Modeling Factors as Taken into Account by GAUSS2D and GELFIT2D} 

\tablehead{\colhead{No.} & \colhead{Factor} & \colhead{GAUSS2D} & \colhead{GELFIT2D}}

\startdata

1. & Spatial distribution & not needed & assumed infinitely thin exponential \\
 & of observable emission & & with possible central hole due to \\
 & & & dust (r$_{em}$, r$_{hole}$ are modeled) \\
 & & & \\
2. & Velocity distribution & assumed arctangent & same as GAUSS2D \\
 & of observable emission & (r$_{to}$, V$_{arc}$ are modeled) & \\
 & & & \\
3. & Disk inclination & input value obtained from & same as GAUSS2D \\
 & & GIM2D model of HST image & \\
 & & of galaxy  & \\
 & & & \\
4. & Seeing blurring & input FWHM obtained by fitting & same as GAUSS2D \\
 & & Gaussian function to profile of & \\
 & & mask alignment star & \\
 & & & \\
5. & Instrumental blurring & input FWHM obtained by convolving   & same as GAUSS2D \\
 & & Gaussian function with slit width and  & \\
 & & matching to night sky line profiles & \\
 & & & \\
6. & Slit width & input value known from slit mask design & same as GAUSS2D \\
 & & & \\
7. & Slit PA vs.\ galaxy SMA & input value obtained by comparing & same as GAUSS2D \\
 & & known slit PA to SMA from GIM2D & \\
 & & model of HST image of galaxy & \\
 & & & \\
8. & Slit position w.r.t. & assumes galaxy is centered in slit & subpixel offsets are modeled \\
 & galaxy center & & \\
 & & & \\
9. & Anamorphic factor & input value known from instrument & same as GAUSS2D \\
 & & design and known grating angle & \\
 & & & \\
10. & Pixelization & input value based on known resolution & same as GAUSS2D \\
 & & of observed spectra & \\
 & & & \\
11. & Atmospheric dispersion & not modeled & modeled if several emission lines \\
 & & & fit separately for one galaxy \\
 & & & \\
12. & Disk thickness & not modeled & not modeled \\
 & & & \\
13. & Presence of bulge, & not modeled & not modeled \\
 & bar, warps & & \\

\enddata
\end{deluxetable}

\clearpage

\begin{landscape}

\begin{deluxetable}{lrrrrrrrrrrl} 
\tabletypesize{\scriptsize}
\tablecolumns{10} 
\tablewidth{0pc} 
\tablecaption{Velocity measurements from GAUSS2D and GELFIT2D} 

\tablehead{\colhead{} &
\colhead{} &
\colhead{} &
\colhead{GAUSS2D} &
\colhead{} &
\colhead{} &
\colhead{} &
\colhead{GELFIT2D} &
\colhead{} &
\colhead{} &
\colhead{FINAL} &
\colhead{}\\  
 \cline{3-5}  \cline{7-9} \cline{11-11}
\colhead{object} &
\colhead{$\Delta$PA} & 
\colhead{$r_{to}$} & 
\colhead{$V_{arc}$} & 
\colhead{$\chi^{2}$} & 
\colhead{} & 
\colhead{$r_{to}$} & 
\colhead{$V_{arc}$} & 
\colhead{$\chi^{2}$} & 
\colhead{} & 
\colhead{$V_{arc}$} & 
\colhead{Lines measured } \\
\colhead{} & 
\colhead{($degrees$)} & 
\colhead{($arcsec$)} & 
\colhead{($km\ s^{-1}$)} & 
\colhead{} & 
\colhead{} & 
\colhead{($arcsec$)} & 
\colhead{($km\ s^{-1}$)} & 
\colhead{} & 
\colhead{} & 
\colhead{($km\ s^{-1}$)} & 
\colhead{($rest$ \AA)} \\
\colhead{(1)} & 
\colhead{(2)} & 
\colhead{(3)} & 
\colhead{(4)} & 
\colhead{(5)} & 
\colhead{} & 
\colhead{(6)} & 
\colhead{(7)} & 
\colhead{(8)} & 
\colhead{} & 
\colhead{(9)} & 
\colhead{(10)} }  
\startdata 
TFR 01	 &  --10	& $0.10^{+0.05}_{-0.00}$	 &  $-240^{+20}_{-45}$	& 2.28 	 & \ &  $\cdots$	 		 &  $\cdots$	 	 &  $\cdots$	 	 & \ &  $-240^{+20}_{-45}$ &  5007,4959,4861,3869,3727,3426 \\
TFR 02	 &  -29	& $0.10^{+0.30}_{-0.00}$	 &  $-120^{+35}_{-45}$	& 1.09 	& \ &  $0.55^{+0.25}_{-0.15}$	 &  $-255^{+55}_{-65}$ 	 & 1.11	& \ &  $-160^{+75}_{-75}$ &  6583,6563,4861 \\
TFR 03	 &  --20	& $0.20^{+0.20}_{-0.10}$	 &  $220^{+30}_{-25}$	 & 1.10	 & \ & $0.35^{+0.00}_{-0.01}$	 &  $225^{+10}_{-15}$	 & 1.65 &  \ & $225^{+10}_{-15}$	 &  6583,6563 \\
TFR 04	 &  --13	& $0.03^{+0.10}_{-0.00}$	 &  $-120^{+20}_{-10}$	& 1.42 & \	 &  $0.09^{+0.08}_{-0.07}$	 &  $-120^{+5}_{-15}$	 & 1.26 & \ & $-120^{+5}_{-10}$	 &  6583,6563 \\
TFR 04	 &  --8	& $0.05^{+0.10}_{-0.00}$	 &  $-90^{+20}_{-35}$	 	 &  2.22 & \ & $0.01^{+0.00}_{-0.00}$	 &  $-110^{+5}_{-15}$	 & 1.39 &  \ & $-110^{+10}_{-20}$ &  6583,6563,5007,4861,4340,3727 \\
TFR 05	 &  +1	& $0.50^{+0.20}_{-0.15}$	 &  $-320^{+55}_{-25}$	 & 1.22	 & \ &  $0.39^{+0.17}_{-0.08}$	 &  $-295^{+20}_{-45}$ 	 & 1.23 &  \ & $-305^{+25}_{-25}$ &  6583,6563,4861 \\
TFR 06	 &  --26	& $0.10^{+0.30}_{-0.00}$	 &  $120^{+45}_{-30}$	 	 & 1.09 & \ & $0.40^{+0.22}_{-0.21}$	 &  $135^{+35}_{-15}$	 & 1.27 &  \ & $130^{+30}_{-15}$	 &  6583,6563,4861 \\
TFR 07	 &  --20	& $0.30^{+0.05}_{-0.15}$	 &  $100^{+5}_{-10}$	 	 & 1.48 & \ & $0.01^{+0.00}_{-0.00}$	 &  $105^{+5}_{-5}$	 & 1.41 & \ & $105^{+5}_{-5}$	 &  6717,6583,6563,5007,4959,4861,4340 \\
TFR 08	 &  +60	& $0.10^{+0.05}_{-0.00}$	 &  $-110^{+20}_{-20}$	 & 2.06	& \ &  $0.02^{+0.01}_{-0.00}$	 &  $-145^{+15}_{-10}$	 & 2.08 & \ & $-135^{+20}_{-20}$ &  5007,4861,4340 \\
TFR 09	 &  +41	& $0.10^{+0.20}_{-0.00}$	 &  $180^{+20}_{-40}$	 	 &  1.36 & \ & $0.02^{+0.06}_{-0.01}$	 &  $150^{+15}_{-20}$	 & 1.16 & \ & $160^{+20}_{-25}$	 &  6583,6563,5007,4959,4861 \\
TFR 10	 &  +17	& $0.40^{+0.10}_{-0.05}$	 &  $-375^{+15}_{-15}$	 & 3.85  & \ &  $0.50^{+0.03}_{-0.16}$	 &  $-365^{+20}_{-25}$	 & 1.71 & \ & $-370^{+15}_{-15}$ &  6583,6563 \\
TFR 11	 &  +2	& $0.30^{+0.15}_{-0.20}$	 &  $260^{+25}_{-45}$	  	 & 1.19 & \ & $0.28^{+0.07}_{-0.07}$	 &  $240^{+25}_{-20}$	 & 1.23 & \ &  $245^{+20}_{-20}$	 &  6583,6563,4861 \\
TFR 12	 &  +16	& $0.20^{+0.30}_{-0.10}$	 &  $-200^{+40}_{-50}$	 & 1.50	 & \ &  $0.01^{+0.01}_{-0.00}$	 &  $-180^{+15}_{-30}$	 & 1.16 & \ & $-185^{+15}_{-30}$ &  6583,6563 \\
TFR 13	 &  +2	& $\cdots$	 	 &  $\cdots$	 	 	 &  $\cdots$	 	& \  	 &  $0.51^{+0.16}_{-0.15}$	 &  $190^{+50}_{-30}$	 & 1.16 & \ & $190^{+50}_{-30}$	 &  6583,6563 \\
TFR 14	 &  --55	& $0.30^{+0.20}_{-0.15}$	 &  $200^{+45}_{-30}$	 	 & 1.06 & \ & $0.02^{+0.02}_{-0.01}$	 &  $170^{+20}_{-15}$	 & 1.29 & \ & $175^{+25}_{-20}$	 &  6563,5007,4861 \\
TFR 15	 &  +43	& $0.10^{+0.05}_{-0.00}$	 &  $-175^{+15}_{-25}$	 & 2.08	 & \ &  $0.01^{+0.01}_{-0.00}$	 &  $-170^{+15}_{-15}$	 & 1.75 & \ & $-170^{+10}_{-15}$	 &  6583,6563,5007,4861,4340 \\
TFR 15	 &  +52	& $0.10^{+0.15}_{-0.00}$	 &  $-140^{+25}_{-30}$	 & 1.32	& \ &  $0.01^{+0.00}_{-0.00}$	 &  $-200^{+15}_{-20}$	 &  1.95 & \ & $-185^{+35}_{-35}$ &  6583,6563,5007,4861 \\
\enddata 
\tablecomments{Columns: (1) Object identification number; (2) offset of slit position angle from galaxy major axis (note duplicate observations with different $\Delta$PA for TFR 04 and TFR 15); (3) turnover radius with 1$\sigma$ errors as measured with GAUSS2D; (4) rotation velocity with 1$\sigma$ errors as measured with GAUSS2D; (5) reduced chi-square of best-fit GAUSS2D model; (6) turnover radius with 1$\sigma$ errors as measured with GELFIT2D; (7) rotation velocity with 1$\sigma$ errors as measured with GELFIT2D; (8) reduced chi-square of best-fit GELFIT2D model; (9) weighted mean of rotation velocities from GAUSS2D and GELFIT2D with 1$\sigma$ errors (10) Rest-frame emission lines measured (6717 = [\ion{S}{2}], 6583 = [\ion{N}{2}], 6563 = H$\alpha$, 5007 = [\ion{O}{3}], 4959 = [\ion{O}{3}], 4861 = H$\beta$, 4340 = H$\gamma$, 3869 = [\ion{Ne}{3}], 3727 = [\ion{O}{2}] doublet, 3426 = [\ion{Ne}{5}].}

\end{deluxetable}
\clearpage
\end{landscape} 

\begin{deluxetable}{rll} 
\tablecolumns{3} 
\tablewidth{0pc} 
\tablecaption{Converting $V_{arc}$ to Locally-Used Velocity Widths} 

\tablehead{\colhead{No.} & \colhead{Conversion} & \colhead{Method}}

\startdata

1. & $V_{arc}^{i} * {\rm sin \ } i = V_{arc}$ & take out inclination correction \\
2. & $V_{arc} \rightarrow V_{\rm pmm}$ & calibrated by analyzing data in Courteau 1997 \\
3. & $V_{\rm pmm} \rightarrow W_{50}$ & using equation B6 in KFF02 \\
4. & $W_{50}$ / sin $i$ $= W_{V_{\rm pmm}}$ & inclination correction \\
 & & to compare to KFF02 \\
 & & \\
5. & $W_{50} + 20$ km s$^{-1} = W_{20}$ & from KFF02 and Haynes et al.\ 1999 \\
6. & $W_{20}$ / sin $i$ $= W_{20}^{i}$ & inclination correction \\
7. & $W_{20}^{i} \rightarrow W_{R}^{i}$ & using Tully \& Fouqu\'{e} 1985 turbulence correction \\
 & & to compare to PT92 and TP00 \\

\enddata

\end{deluxetable}

\clearpage

\begin{landscape}
\begin{deluxetable}{lrrrrrrrrrrrrrrr} 
\tabletypesize{\scriptsize}
\tablecolumns{16} 
\tablewidth{0pt} 
\tablecaption{Color, Luminosity, and Velocity Measurements} 

\tablehead{
\colhead{} & \colhead{} & \ &
\multicolumn{2}{c}{``raw'' measurements} & \ &
\multicolumn{2}{c}{compare to PT92} & \ &
\multicolumn{2}{c}{compare to TP00} & \ &
\multicolumn{2}{c}{compare to KFF02} & \ & \colhead{} \\
\cline{4-5} \cline{7-8} \cline{10-11} \cline{13-14} \\
\colhead{object} & 
\colhead{$(U-B)_{0}$} & \ &
\colhead{$M_{B}^{\rm dist}$} & \colhead{$2*\vert V_{arc}\vert$} & \ &
\colhead{$M_{B}^{\rm PT92}$} & \colhead{$W_{R}^{i}$} & \ &
\colhead{$M_{B}^{\rm TP00}$} & \colhead{$W_{R}^{i}$} & \ &
\colhead{$M_{B}^{\rm KFF02}$} & \colhead{$W_{V_{\rm pmm}}$} &
\ & \colhead{vel.\ err.} \\
\colhead{} & \colhead{({\it mag})} & \ & 
\colhead{({\it mag})} & \colhead{({\it km s}$^{-1}$)} & \ &
\colhead{({\it mag})} & \colhead{({\it km s}$^{-1}$)} & \ &
\colhead{({\it mag})} & \colhead{({\it km s}$^{-1}$)} & \ &
\colhead{({\it mag})} & \colhead{({\it km s}$^{-1}$)} & 
\ & \colhead{({\it km s}$^{-1}$)}\\
\colhead{(1)} & \colhead{(2)} & \ &
\colhead{(3)} & \colhead{(4)} & \ &
\colhead{(5)} & \colhead{(6)} & \ & 
\colhead{(7)} & \colhead{(8)} & \ &
\colhead{(9)} & \colhead{(10)} & \ &
\colhead{(11)} }
\startdata
TFR01 & $\cdots$ & & -21.41 & 480 & & -21.35 & 545 & & -21.79 & 545 & & -21.73 & 560 & \ & $^{+40}_{-90}$ \\
TFR02 & $\cdots$ & & -20.45 & 320 & & -20.39 & 375 & & -20.29 & 375 & & -20.23 & 380 & \ & $^{+150}_{-150}$ \\
TFR03 &     0.29 & & -20.67 & 450 & & -20.62 & 510 & & -21.25 & 510 & & -21.20 & 525 & \ & $^{+20}_{-30}$ \\
TFR04 &    -0.34 & & -21.10 & 240 & & -21.03 & 265 & & -21.17 & 265 & & -21.09 & 285 & \ & $^{+10}_{-20}$ \\
TFR05 &     0.18 & & -21.09 & 610 & & -21.04 & 695 & & -21.31 & 695 & & -21.26 & 710 & \ & $^{+50}_{-50}$ \\
TFR06 &     0.14 & & -20.55 & 260 & & -20.50 & 295 & & -20.46 & 295 & & -20.41 & 305 & \ & $^{+60}_{-30}$ \\
TFR07 &    -0.22 & & -20.50 & 210 & & -20.43 & 250 & & -20.32 & 250 & & -20.24 & 250 & \ & $^{+10}_{-10}$ \\
TFR08 &    -0.23 & & -18.05 & 270 & & -17.99 & 310 & & -17.96 & 310 & & -17.90 & 320 & \ & $^{+40}_{-40}$ \\
TFR09 &    -0.11 & & -21.00 & 320 & & -20.94 & 375 & & -20.85 & 375 & & -20.79 & 380 & \ & $^{+40}_{-50}$ \\
TFR10 &     0.06 & & -22.24 & 740 & & -22.20 & 840 & & -22.80 & 840 & & -22.76 & 860 & \ & $^{+30}_{-30}$ \\
TFR11 &     0.30 & & -20.85 & 490 & & -20.80 & 555 & & -21.35 & 555 & & -21.30 & 570 & \ & $^{+40}_{-40}$ \\
TFR12 &     0.10 & & -21.11 & 370 & & -21.05 & 440 & & -20.94 & 440 & & -20.87 & 440 & \ & $^{+30}_{-60}$ \\
TFR13 &     0.32 & & -20.05 & 380 & & -20.00 & 435 & & -20.08 & 435 & & -20.02 & 445 & \ & $^{+100}_{-60}$ \\
TFR14 &    -0.02 & & -19.88 & 350 & & -19.82 & 405 & & -19.80 & 405 & & -19.72 & 410 & \ & $^{+50}_{-40}$ \\
TFR15 &    -0.01 & & -21.09 & 340 & & -21.03 & 395 & & -20.95 & 395 & & -20.88 & 400 & \ & $^{+20}_{-30}$ \\
\enddata
\tablecomments{Columns: (1) object identification number; (2) restframe $U - B$ aperture color, calculated using BH82 Galactic extinction and $A_{U}^{\rm T98,2}$ and $A_{B}^{\rm T98,2}$ internal extinction corrections; (3) absolute restframe $B$ magnitude, SFD98 Galactic and TF85 internal extinction corrections applied; (4) arctangent rotation velocity width; (5) absolute restframe $B$ magnitude, calculated using BH82 Galactic and TF85 internal extinction corrections; (6) rotation velocity width derived using steps 1 -- 7 of Table 6; (7) absolute restframe $B$ magnitude, SFD98 Galactic and T98,1 internal extinction corrections applied; (8) rotation velocity width derived using steps 1 - 7 of Table 6; (9) absolute restframe $B$ magnitude, BH82 Galactic and T98,2 internal extinction corrections applied; (10) ``probable min-max'' velocity width derived using steps 1 -- 4 of Table 6; (11) error on velocity measurement, twice that derived from the arctangent velocity model.}
\end{deluxetable}
\clearpage
\end{landscape}


\begin{center}
\begin{deluxetable}{lrrrr} 
\tabletypesize{\scriptsize}
\tablecolumns{5} 
\tablewidth{0pt} 
\tablecaption{Internal Extinction Corrections} 

\tablehead{
\colhead{object} & 
\colhead{$A_{B}^{\rm TF85}$} & 
\colhead{$A_{B}^{\rm T98,1}$} & 
\colhead{$A_{U}^{\rm T98,2}$} & 
\colhead{$A_{B}^{\rm T98,2}$} \\ 
\colhead{} & 
\colhead{({\it mag})} & 
\colhead{({\it mag})} & 
\colhead{({\it mag})} & 
\colhead{({\it mag})} \\
\colhead{(1)} & \colhead{(2)} & 
\colhead{(3)} & \colhead{(4)} & 
\colhead{(5)} }
\startdata
TFR01 & 0.65 & 1.03 & 1.11 & 1.03  \\
TFR02 & 0.33 & 0.17 & 0.18 & 0.17  \\
TFR03 & 0.88 & 1.46 & 1.57 & 1.46  \\
TFR04 & 0.92 & 0.99 & 1.06 & 0.98  \\
TFR05 & 0.47 & 0.69 & 0.75 & 0.69  \\
TFR06 & 0.43 & 0.34 & 0.36 & 0.34  \\
TFR07 & 0.35 & 0.17 & 0.17 & 0.16  \\
TFR08 & 0.41 & 0.32 & 0.34 & 0.32  \\
TFR09 & 0.34 & 0.19 & 0.20 & 0.19  \\
TFR10 & 0.61 & 1.17 & 1.26 & 1.17  \\
TFR11 & 0.75 & 1.25 & 1.35 & 1.25  \\
TFR12 & 0.31 & 0.14 & 0.15 & 0.13  \\
TFR13 & 0.44 & 0.47 & 0.50 & 0.46  \\
TFR14 & 0.37 & 0.28 & 0.30 & 0.27  \\
TFR15 & 0.34 & 0.20 & 0.21 & 0.19  \\
\enddata
\tablecomments{Columns: (1) object identification number; (2) inclination-dependent $B$-band internal extinction correction from TF85; (3) inclination- and velocity-dependent $B$-band internal extinction correction from T98, where velocities used for correction were calculated according to steps 1 -- 7 of Table 6; (4) $U$-band internal extinction correction from T98, where velocities used for correction were calculated according to Table 6 but with steps 6 and 7 reversed, as in KFF02; (5) $B$-band internal extinction correction from T98, where velocities used for correction were calculated according to Table 6 but with steps 6 and 7 reversed, as in KFF02.} 
\end{deluxetable}
\end{center}



\begin{deluxetable}{lrrrrrr} 
\tablecolumns{6} 
\tablewidth{0pc} 
\tablecaption{GIM2D and GELFIT2D Residual and Asymmetry Measurements} 

\tablehead{\colhead{} & \colhead{$\Delta$TF} & \colhead{$\Delta$TF} & \colhead{GIM2D} & \colhead{GIM2D} & \colhead{GELFIT2D} & \colhead{GELFIT2D} \\
\colhead{Object} & \colhead{({\it mag})} & \colhead{({\it km s$^{-1}$})} & \colhead{Residual\tablenotemark{a}} & \colhead{Asymmetry\tablenotemark{b}} & \colhead{Residual} & \colhead{Asymmetry}}

\startdata
TFR 01 & $0.10^{+0.34}_{-0.37}$ & $15^{+25}_{-50}$ & 0.08 & 0.05 & $\cdots$ & $\cdots$ \\
TFR 02 & $-0.092^{+1.47}_{-1.47}$ & $-10^{+75}_{-75}$ & 0.06 & 0.05 & 0.05 & 0.16 \\
TFR 03 &   $0.36^{+0.22}_{-0.27}$ & $40^{+20}_{-25}$ & 0.15 & 0.08 & 0.53 & 0.40 \\
TFR 04 & $-2.22^{+0.21}_{-0.26}$ & $-190^{+15}_{-20}$ & 0.18 & 0.08 & 0.17 & 0.08 \\
TFR 05 &  $1.61^{+0.33}_{-0.37}$ & $220^{+30}_{-30}$ & 0.04 & 0.03 & 0.01 & 0.05 \\
TFR 06 & $-1.24^{+0.80}_{-0.82}$ & $-100^{+35}_{-20}$ & 0.08 & 0.07 & 0.15 & 0.17 \\
TFR 07 & $-1.94^{+0.23}_{-0.27}$ & $-140^{+15}_{-15}$ & 0.13 & 0.09 & 0.15 & 0.07 \\
TFR 08 &   $1.48^{+0.54}_{-0.56}$ & $90^{+20}_{-20}$ & 0.06 & 0.06 & 0.15 & 0.09 \\
TFR 09 & $-0.65^{+0.46}_{-0.49}$ & $-60^{+25}_{-30}$ & 0.14 & 0.09 & --0.01 & 0.01 \\
TFR 10 &   $0.96^{+0.21}_{-0.26}$ & $170^{+30}_{-30}$ & 0.13 & 0.06 & 0.27 & 0.26 \\
TFR 11 &   $0.61^{+0.33}_{-0.37}$ & $75^{+25}_{-25}$ & 0.06 & 0.04 & 0.26 & --0.01 \\
TFR 12 & $-0.098^{+0.33}_{-0.36}$ & $-10^{+20}_{-35}$ & 0.09 & 0.05 & 0.12 & 0.01 \\
TFR 13 &   $0.80^{+0.90}_{-0.91}$ & $75^{+50}_{-35}$ & 0.12 & 0.09 & 0.01 & --0.01 \\
TFR 14 &   $0.74^{+0.53}_{-0.55}$ & $65^{+30}_{-25}$ & 0.11 & 0.07 & 0.27 & 0.33 \\
TFR 15 & $-0.52^{+0.26}_{-0.30}$ & $-50^{+20}_{-20}$ & 0.10 & 0.09 & 0.12 & 0.07 \\
\enddata
\tablenotetext{a}{Tran et al.\ (2001, 2003) considered GIM2D residual$=0.1$ to be the cutoff between low- and high-residual objects.}
\tablenotetext{a}{Tran et al.\ (2001, 2003) considered GIM2D asymmetry$=0.05$ to be the cutoff between relatively symmetric and asymmetric objects.}
\end{deluxetable}

\clearpage

\begin{deluxetable}{lrrrrr} 
\tablecolumns{6} 
\tablewidth{0pc} 
\tablecaption{Magnitude Selection Effects on Tully-Fisher Slope Measurements} 

\tablehead{\colhead{} & 
\multicolumn{2}{c}{Tully-Fisher Relation\tablenotemark{a}} & \ & 
\multicolumn{2}{c}{Velocity-$\Delta$TF Relation\tablenotemark{b}} \\
\colhead{} & 
\colhead{} & 
\colhead{} & \ &
\multicolumn{2}{c}{input T-F relation known} \\
\cline{2-3} \cline{5-6} \\
\colhead{Sample} & 
\colhead{Slope} & 
\colhead{Zeropoint} & \ &
\colhead{Slope} & 
\colhead{Zeropoint} }
\startdata
Input relations   & $-10.09$         & $-19.83$        & \ & $0.00$        & $0.00$  \\
Low-dispersion    & $-10.09\pm0.14$  & $-19.83\pm0.03$ & \ & $0.56\pm0.13$ & $-1.41\pm0.21$ \\
Medium-dispersion & $-9.60\pm0.29$   & $-19.87\pm0.06$ & \ & $2.47\pm0.23$ & $-6.29\pm0.36$ \\
High-dispersion   & $-9.16\pm0.65$   & $-20.04\pm0.16$ & \ & $6.85\pm0.23$ & $-17.40\pm0.37$ \\
\enddata
\tablenotetext{a}{We use the following functional form for the Tully-Fisher relation: $M_{B} = {\rm zeropoint} + {\rm slope}[{\rm log}(W_{V_{\rm pmm}}) - 2.5]$ }
\tablenotetext{b}{The functional form of the velocity-$\Delta$TF relation is: $\Delta$TF $= {\rm zeropoint} + {\rm slope}*{\rm log}(W_{V_{\rm pmm}})$}
\end{deluxetable}

\clearpage
\appendix
\section{Notes on Individual Objects}

For ease of comparison to previous works, we provide identification numbers from Schneider, Dressler \& Gunn (1986; SDG), where available, and from Czoske et al.\ (2001; Cz01).  \\

\noindent {\bf TFR 01:} Dropped GELFIT2D models from V$_{arc}$ determination due to very poor fits ($\chi^{2} > 10$).  Emission appears concentrated toward center of galaxy ($r_{d,emission} \simeq 0.2\arcsec$;  $r_{d,image} \simeq 1.1\arcsec$).  Kept GAUSS2D V$_{arc}$ measurement, but note that this may not be representative of velocity distribution of the entire galaxy.  This object exhibits a point source at its center as well as [\ion{Ne}{3}] $\lambda3869$ and [\ion{Ne}{5}] $\lambda3426$ emission, indicative of AGN activity.  Low [\ion{O}{2}] flux ratio (I($\lambda3729$)/I($\lambda3726$) = 0.4) indicates high electron density ($N_{e} \ge 1000$ cm$^{-1}$).  Wavelength range of LRIS spectrum does not cover H$\alpha$ or [\ion{N}{2}].  Only early-type (S0) in TFR sample.  Cross-reference: Cz01 149.\\

\noindent {\bf TFR 02:}  GELFIT2D and GAUSS2D velocity measurements differ most for this object, with $v_{rot}^{GEL} \sim 2v_{rot}^{GAUSS}$. Perhaps uncertainty is due to relatively low inclination.  Object lies at slightly lower redshift that the main cluster; possibly member of a recently infallen subcluster (Czoske et al.\ 2002).  Cross-reference: Cz01 265.\\

\noindent {\bf TFR 03:} Reddest galaxy in TFR sample.  WFPC2 image of object appears smooth and symmetric; however, emission is clearly brighter toward top of slit than toward the bottom: object not centered in slit?  As with TFR 02, this object lies at a slightly low redshift and may reside in a recently infallen subcluster.  Cross-reference: Cz01 216.\\

\noindent {\bf TFR 04:} Very blue object with odd morphology, poorly fit by exponential disk.  Unusual morphology also noted by Lavery, Pierce, \& McClure (1992) may be indicative of tidal distortion.  Poor GAUSS2D and GELFIT2D fits imply that this galaxy does not have a symmetric, circular velocity distribution.  Koo et al.\ (1997) derive a linewidth for this object of 40 km s$^{-1}$; we find $\sigma = 75$ km s$^{-1}$. Cross-references: SDG 231, Cz01 332.\\

\noindent {\bf TFR 05:} Object lies in a relatively dense region of spectroscopically confirmed cluster members (nearest cluster member at distance  3.8$\arcsec$ on sky). Cross-references: SDG 194, Cz01 318.\\

\noindent {\bf TFR 06:} Two nearby galaxies are also spectroscopically confirmed cluster members, possible nearby tidal tails.  Closest companion has distance of 2.8$\arcsec$ on sky. Cross-references: SDG 189, Cz01 434.\\

\noindent {\bf TFR 07:} Clear spiral structure with arms.  One of three bluest galaxies in sample (only TFR 04 and TFR 08 have comparable colors). Cross-references: SDG 175, Cz01 464.\\

\noindent {\bf TFR 08:} Very blue, clumpy object near cluster center; by far the least luminous galaxy in TFR sample.  Odd morphology noted by both Lavery, Pierce, \& McClure (1992) and Treu et al.\ (2003).  Red companion is also spectroscopically confirmed cluster member at a distance of 2.0$\arcsec$ on sky.  Poor GAUSS2D and GELFIT2D fits imply that this galaxy does not have a symmetric, circular velocity distribution.  Object undergoing tidal distortion.  Koo et al.\ (1997) derive a linewidth for this object of 50 km s$^{-1}$; we find $\sigma = 40$ km s$^{-1}$. Cross-references: SDG 173A, Cz01 379.\\

\noindent {\bf TFR 09:} Clear spiral structure with arms, fairly blue.  Velocity distribution appears asymmetric.  Cross-reference: Cz01 466.\\

\noindent {\bf TFR 10:} Largest, most luminous object in sample; one of two reddest galaxies in sample.  F450W image indicates ring of star formation.  Considered ``disturbed" by Lavery, Pierce, \& McClure (1992).  Clumpy GELFIT2D residuals, clear wiggles in measured GAUSS2D velocity curve.  Koo et al.\ (1997) derive a linewidth of 120 km s$^{-1}$ for this object; we find $\sigma = 115$ km s$^{-1}$.  Cross-references: SDG 146, Cz01 459.\\

\noindent {\bf TFR 11:} This large, red disk galaxy has a nearby disk companion (distance = 4.0$\arcsec$ on sky), also a spectroscopically confirmed cluster member.  Cross-references: SDG 129, Cz01 429.\\

\noindent {\bf TFR 12:} Object may have faint companion (off edge of WFPC2 chip; unable to determine redshift from LRIS spectrum).  Cross-references: SDG 121, Cz01 172.\\

\noindent {\bf TFR 13:} Dropped GAUSS2D model from V$_{arc}$ determination as low S/N emission yielded few velocity curve measurements.  However, GELFIT2D fits to two emission lines yield consistent V$_{arc}$ measurements.  One of the reddest objects in the TFR sample.  Cross-references: SDG 119, Cz01 396.\\

\noindent {\bf TFR 14:} Large $\Delta$PA between slit and semi-major axis for this object.  Cross-references: SDG 72, Cz01 238.\\

\noindent {\bf TFR 15:} This galaxy lies near two other confirmed cluster members, at distances 3.4$\arcsec$ and 4.2$\arcsec$ on sky.  Treu et al.\ (2003) designate this object as a barred galaxy. Cross-reference: Cz01 199.\\

\end{document}